\documentclass[iop, revtex4,pdflatex]{emulateapj}
\usepackage{times}
\usepackage{graphicx}
\usepackage{ctable}
\usepackage{romannum}
\usepackage{array}
\pdfoutput=1 
\usepackage[backref,breaklinks,colorlinks,citecolor=blue]{hyperref}
\usepackage[all]{hypcap}
\usepackage{multirow}
\usepackage{array}
\usepackage{appendix}
\usepackage{booktabs}
\usepackage{multirow}
\newcolumntype{+}{>{\global\let\currentrowstyle\relax}}
\newcolumntype{^}{>{\currentrowstyle}}

\newcommand{\sw}{\textit{Swift}}
\newcommand{\ha}{H{\small $\alpha$}}
\newcommand{\pb}{Pa{\small $\beta$}}
\newcommand{\hb}{H{\small $\beta$}}

\def\cygx1{Cygnus~X$-$1}
\def\gx{GX~339$-$4}

\def\swj{Swift J1753.5$-$0127}

\def\chis{$\chi^2$}

\def\msun{$M_{\odot}$}
\def\rsun{$R_{\odot}$}

\def\ergcms{erg~cm$^{-2}$~s$^{-1}$}
\def\kms{km~s$^{-1}$}

\def\wm2{W~m$^{-2}$}
\def\mic{$\mu$m}
\def\cm2{cm$^{-2}$}
\def\se1{s$^{-1}$}
\def\nh{$N_{\rm H}$}
\def\nhe{N_{\rm H}}

\def\Ave{A_{\rm V}}

\begin{document}

\title{Optical and near-infrared spectroscopy of the black hole Swift
  J1753.5-0127}
\thanks{Based on observations performed with European Southern
Observatory (ESO) Telescopes at the Paranal Observatory under programmes ID 093.D-0786}

\author{Farid Rahoui\altaffilmark{1,2}, John
  A. Tomsick\altaffilmark{3}, Mickael Coriat\altaffilmark{4},
                                St{\'e}phane Corbel\altaffilmark{5,6},
                                Felix F{\"u}rst\altaffilmark{7},
                                Poshak Gandhi\altaffilmark{8}, Emrah
                                Kalemci\altaffilmark{9}, Simone
                                Migliari\altaffilmark{10}, Daniel
                                Stern\altaffilmark{11}, Anastasios
                                K. Tzioumis\altaffilmark{12}
}
\email{frahoui@eso.org}

\altaffiltext{1}{European Southern Observatory,
  K. Schwarzschild-Str. 2, 85748 Garching bei M\"unchen, Germany} 
\altaffiltext{2}{Department of Astronomy, Harvard University, 60
  Garden street, Cambridge, MA 02138, USA}
\altaffiltext{3}{Space Sciences Laboratory, 7 Gauss Way, University of
  California, Berkeley, CA 94720-7450, USA}
\altaffiltext{4}{IRAP, Universit{\'e} de Toulouse, UPS, 9 Avenue du
  colonel Roche, F-31028 Toulouse Cedex 4, France; CNRS, UMR5277,
  F-31028 Toulouse, France}
\altaffiltext{5}{Laboratoire AIM (CEA/IRFU - CNRS/INSU -
  Universit{\'e} Paris Diderot), CEA DSM/IRFU/SAp, F-91191
  Gif-sur-Yvette, France} 
\altaffiltext{6}{Station de Radioastronomie de Nançay, Observatoire de
  Paris, PSL Research University, CNRS, Univ. Orléans, OSUC, 18330
  Nançay, France}
\altaffiltext{7}{California Institute of Technology, 1200 East
  California Boulevard, Pasadena, CA 91125, USA}
\altaffiltext{8}{Department of Physics, University of Durham, South
  Road, Durham DH1 3LE, UK}
\altaffiltext{9}{Sabanci University, Orhanli-Tuzla, Istanbul, 34956,
  Turkey} 
\altaffiltext{10}{European Space Astronomy Centre, Apartado/P.O. Box
  78, Villanueva de la Canada, E-28691 Madrid, Spain}
\altaffiltext{11}{Jet Propulsion Laboratory, California Institute of
  Technology, 4800 Oak Grove Drive, Pasadena, CA 91109, USA}
\altaffiltext{12}{CSIRO Astronomy and Space Science, Australia
  Telescope National Facility, P.O. Box 76, Epping, NSW 1710,
  Australia}

\shorttitle{X-shooter spectroscopy of \swj}
\shortauthors{F. Rahoui et al.}
\submitted{{\sc Accepted to ApJ:} August 3, 2015}
\begin{abstract}

We report on a multiwavelength observational campaign of the
  black hole X-ray binary \swj\ that consists of an ESO/X-shooter
  spectrum supported by contemporaneous \sw/XRT+UVOT and ATCA
  data. ISM absorption lines in the X-shooter spectrum allows us to
  determine $E(B-V)=0.45\pm0.02$ along the line-of-sight to the
  source. We also report detection of emission signatures of
  \ion{He}{2}~$\lambda4686$, H{\small $\alpha$},  and, for the first
  time, \ion{H}{1}~$\lambda10906$ and Pa{\small $\beta$}. The
  double-peaked morphology of these four lines is typical of the
  chromosphere of a rotating accretion disk. Nonetheless, the paucity
  of disk features points towards a low level of irradiation in the
system. This is confirmed through spectral energy distribution
modeling and we find that the UVOT+X-shooter continuum mostly stems
from the thermal emission of a viscous disk. We speculate that the
absence of reprocessing is due to the compactness of an
illumination-induced envelope that fails to reflect enough incoming
hard X-ray photons back to the outer regions. The disk also marginally
contributes to the Compton-dominated X-ray emission and is strongly
truncated, with an inner radius about a thousand times larger than the
black hole's gravitational radius. A near-infrared excess is present,
and we associate it with synchrotron radiation from a compact
jet. However, the measured X-ray flux is significantly higher 
than what can be explained by the optically thin synchrotron jet
component. We discuss these findings in the framework of the radio
quiet versus X-ray bright hypothesis, favoring the presence of a
residual disk, predicted by evaporation models, that contributes to
the X-ray emission without enhancing the radio flux.
\end{abstract}


\section{Introduction}

Microquasars are disk-accreting X-ray binaries characterized by
  the presence of collimated bipolar radio ejections called jets. When
  in outburst, they transition between two canonical spectral states,
  dubbed soft when the disk dominates the X-ray emission and hard when
  continuous compact radio jets are present \citep[see
  e.g.][]{2004Fenderb}. In the hard state, microquasars have 
  consequently long been known to exhibit a very tight radio/X-ray
correlation spanning several orders of magnitudes. First measured for
a couple of sources almost 20 years ago \citep{1998Hannikainen,
  2000Corbel}, this relation was progressively extended to almost all
known systems \citep[see e.g.][]{2003Gallo, 2013Corbel,
  2014Gallo}. Nonetheless, it quickly appeared that what was initially
thought to be a universal relation, $F_{\rm R}\propto F_{\rm
  X}^{0.6-0.7}$, is actually not followed by a growing number of
outliers that exhibit a lower-than-expected radio flux for a given
X-ray luminosity \citep[see e.g.][]{2004Corbelb, 2007Gallob,
  2011Coriat}. Such sources are often referred to as radio quiet
microquasars \citep{2011Soleri, 2011Coriat}, by analogy with active
galactic nuclei (AGN).

\begin{table*}
\caption{\scriptsize Summary of the observations of \swj\ we made
  use of in this study.\label{logobs}}
\begin{center}
\begin{tabular}{c|c|cccccc|ccc|cc}
\hline
&\sw/XRT&\multicolumn{6}{c|}{\sw/UVOT}&\multicolumn{3}{c|}{X-shooter}&\multicolumn{2}{c}{ATCA}\\
\hline
\hline
Date\tmark[a]&56885.3&\multicolumn{6}{c|}{56885.3}&\multicolumn{3}{c|}{56886.0}&\multicolumn{2}{c}{56885.3}\\
Band\tmark[b]&0.4--7.5 keV&$uw2$&$um2$&$uw1$&$u$&$b$&$v$&UVB&VIS&NIR&5.5 GHz&9 GHz\\
Exposure\tmark[c]&1920&628&454&304&156&157&156&\multicolumn{3}{c|}{6720}&\multicolumn{2}{c}{6060}\\
\hline
\end{tabular}
\end{center}
\tablenotetext{a}{\scriptsize Starting date of the observation, in MJD}
\tablenotetext{b}{\scriptsize Energy band, filters, or central frequencies}
\tablenotetext{c}{\scriptsize Total exposure time on-source, in seconds}
\end{table*}

\begin{table}
\begin{center}
\caption{\scriptsize \swj\ UVOT flux densities\label{uvot}}
\begin{tabular}{ccc}
\tableline\tableline
Filter\tmark[a]&$\lambda_{\rm c}\tmark[b]$&$F_{\lambda}\tmark[c]$\\
\tableline
{\it uvw2}&1928&8.17$\pm$0.39\\
{\it um2}&2246&6.05$\pm$0.38\\
{\it uw1}&2600&8.55$\pm$0.49\\
{\it u}&3465&10.8$\pm$0.5\\
{\it b}&4392&9.27$\pm$0.42\\
{\it v}&5468&7.81$\pm$0.52\\
\tableline
\end{tabular}
\end{center}
\tablenotetext{a}{\scriptsize Filter designation}
\tablenotetext{b}{\scriptsize Central wavelength in \AA}
\tablenotetext{c}{\scriptsize Flux density in units of $10^{-16}$~\ergcms~\AA$^{-1}$}
\end{table}

One such radio quiet microquasar, \swj\ was discovered by the
\sw\ Burst Alert Telescope on 2005 June 30 \citep{2005Palmer}. It is
the microquasar with the shortest orbital period, first found
  to be lower than 3.25~hours \citep{2008Zuritab}, and later refined
to 2.85~hours \citep{2014Neustroev}. With such a short period, it
is very likely that the companion star is a cold M dwarf, 
  although it has never been formally identified. In
  fact, besides the orbit, very little is known about the parameters
of the system. The presence of non P-Cygni emission lines in its
ultraviolet (UV) spectrum and the absence of eclipse put the
inclination in the range 40\degr$-$80\degr\ \citep{2014Froning} and 
\citet{2009Reis} measured $i={55^{\circ}}_{-7}^{+2}$ from X-ray
spectral fitting. The distance is also poorly constrained but is
believed to be between 1~kpc and 8~kpc \citep{2007Cadolle,
  2008Durant}, although larger values cannot be ruled out. Likewise,
the black hole (BH) mass is unknown, with estimates ranging from a 
typical $M_{\rm BH}\sim10$~\msun\ value \citep{2007Cadolle,
  2008Zuritab}  to $M_{\rm BH}\le5$~\msun\ \citep{2014Neustroev}.

\swj\ has never returned to quiescence since the beginning of its
outburst in 2005, and it remained almost ten years in the hard
  state before transitioning to the soft state in February
  2015. Although this extended period of heightened flux has put
the  source under a thorough scrutiny, the reasons behind this
behavior remain uncertain. Moreover, whether the accretion disk
is truncated as expected in the framework of advection-dominated
accretion flows \citep[ADAF, see e.g.][]{1995Narayan} or extends up to
its innermost stable circular orbit (ISCO) is still unclear. Recently,
\citet{2014Froning} modeled the UV to near-infrared (near-IR) spectral
energy distribution (SED) of the source with a viscous accretion disk
and concluded that the disk had to be strongly truncated. On the other
hand, \citet{2009Reis} fitted the {\it XMM-Newton} spectrum with a
reflection model and found no truncation at all. The compact
jet contribution to the optical and near-IR domains is also unknown. 

In this paper we report on a multiwavelength study of \swj\ in the
hard state, centered on its optical and near-IR spectroscopic emission
as observed in August 2014. Our main goals are (1) to study the geometry
of the system; and (2) to constrain the contribution of the compact jet to
the optical and near-IR emission. A work focusing on its early-April 2014
radio and X-ray properties, when the source was in a low-luminosity
hard state, is presented in a companion paper
\citep[][hereafter T15]{2015Tomsick}.  The observations and data
reduction procedures are presented in Section~2. Section~3 is devoted to
analysis of the optical and near-IR spectrum, and Section~4
focuses on modeling the radio to X-ray SED. We discuss our
results and their implications on our understanding of outliers to the
radio/X-ray correlation of microquasars in Section~5, and we conclude
in Section~6.

\begin{table}
\begin{center}
\caption{\scriptsize \swj\ {\it WISE} flux densities\label{wiseflux}}
\begin{tabular}{ccc}
\tableline\tableline
Observation date&$F_{\rm W1}\tmark[a]$&$F_{\rm W2}\tmark[b]$\\
\tableline
MJD~55274.4&$0.63\pm0.01$&$0.53\pm0.02$\\
MJD~55454.0&$0.53\pm0.02$&$0.44\pm0.02$\\
MJD~55456.5&$0.53\pm0.02$&$0.44\pm0.03$\\
MJD~55736.4&$0.51\pm0.01$&$0.49\pm0.02$\\
MJD~55918.7&$0.51\pm0.01$&$0.43\pm0.02$\\
\tableline
\end{tabular}
\end{center}
\tablenotetext{a}{\scriptsize W1 flux density in mJy}
\tablenotetext{b}{\scriptsize W2 flux density in mJy}
\end{table}

\section{Observations}

The data set consists of contemporaneous observations obtained on 2014
August 16 with (1) ESO+VLT/X-shooter (Obs. ID 093.D-0786, PI Rahoui); (2) the
X-Ray Telescope \citep[XRT,][]{2005Burrows} and the
Ultra-Violet/Optical Telescope \citep[UVOT,][]{2005Roming} mounted on
the \sw\ \citep{2004Gehrels} satellite (Obs.~ID 00033140032, PI Tomsick);
and the Australia Telescope Compact Array (ATCA) at 5.5~GHz and 9~GHz
(PI Coriat). A summary is given in \autoref{logobs}.

\subsection{X-shooter observations}

We obtained medium-resolution spectroscopy of \swj\ with the
three available arms of X-shooter \citep{2011Vernet}, UVB, VIS
and NIR (1\farcs3, 1\farcs2, and 1\farcs2 slit-widths, respectively),
giving simultaneous $300-2480$~nm spectral coverage. Atmospheric
conditions were medium-to-good, with a clear sky, seeing at 500~nm in
the range 1\farcs2-1\farcs5, and an airmass between 1.1 and 1.3. The
exposure  time of each individual frame was set to 210~s, 210~s, and
70~s in the UVB, VIS, and NIR arms, respectively, and a total of 32,
32, and 96 frames were taken, with three near-IR exposures being
  obtained   for each UV/optical exposure. Standard ABBA dithering
was used for effective background subtraction, and the A0V telluric
standard star HR6572 was observed in similar conditions for telluric
features removal and flux-calibration. 

We reduced the data using the dedicated pipeline (v.~2.5.2)
implemented in the ESO data reduction environment {\tt
  Reflex}~v.~2.6 \citep{2013Freudling}. It follows the
standard steps for echelle spectroscopy reduction and produces
a cleaned, background-subtracted, and wavelength-calibrated 2D
spectroscopic image. We then used the routines {\tt apall} and {\tt
  telluric} implemented in {\tt IRAF}~v.~2.16\footnote{ {\tt
      IRAF} is  distributed by the National Optical Astronomy
    Observatories, which are operated by the Association of
    Universities for Research in Astronomy, Inc., under cooperative
    agreement with the National Science Foundation.} to extract the
1D spectra and remove the telluric features. We finally performed 
the flux-calibration following the procedure presented in
\citet{2012Rahoui, 2014Rahouib}, and rebinned the spectra from
their original pixel resolution, i.e. 0.2~\AA, 0.2~\AA, and
0.6~\AA, to 1~\AA, 1~\AA, and 3~\AA\ in the UVB, VIS, and NIR arms,
respectively.

\subsection{\sw\ observations}

\begin{figure}[h!]
\begin{center}
\includegraphics[width=9cm]{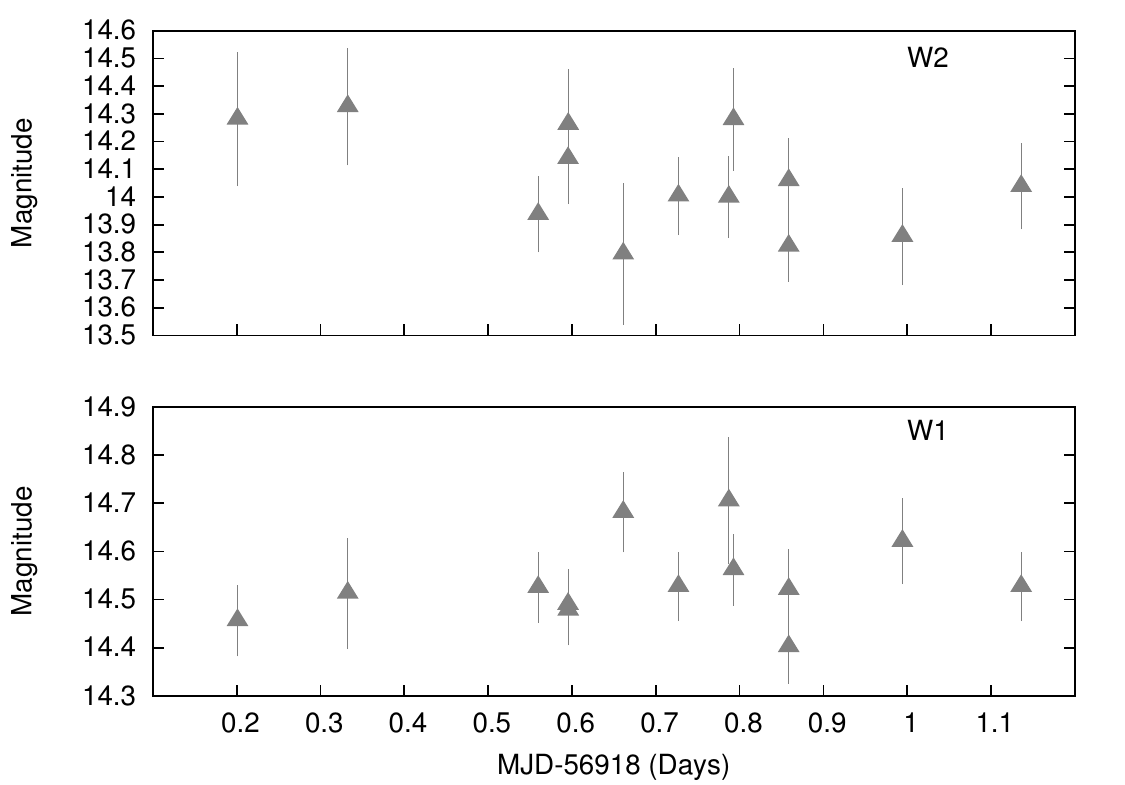}
\caption{\small \swj\ {\it WISE} magnitudes in W1 (3.35~\mic, bottom) and W2
  (4.6~\mic, top) obtained about one month after our observations at
  13 different epochs. No variability is present at 2-$\sigma$.} 
\label{wise}
\end{center}
\end{figure}

\begin{table*}
\begin{center}
\caption{\scriptsize Optical and near-IR lines in the \swj\ spectrum.\label{speclines}}
\begin{tabular}{l|cc|c|cc|c|c}
\tableline\tableline
&\multicolumn{2}{c|}{\ion{He}{2}}&H{$\beta$}&\multicolumn{2}{c|}{H{$\alpha$}}&\ion{He}{1}&Pa{$\beta$}\\
\tableline
${\lambda_{\rm \small c}}^a$&4668$\pm$2&4704$\pm$4&4859$\pm$7&6547$\pm$2&6580$\pm$2&10906$\pm$9&12815$\pm$6\\
${\mathring{W}}^b$&-2.3$\pm$0.5&-1.7$\pm$0.4&+4.1$\pm$1.4&-1.4$\pm$0.2&-1.5$\pm$0.3&-8.8$\pm1.5$&-7.5$\pm$1.2\\
FWHM$^c$&1948$\pm$199&1481$\pm$183&8762$\pm$1147&1244$\pm$174&1250$\pm$111&3771$\pm$474&2904$\pm$319\\
${F_{\rm line}}^d$&1.95$\pm0.44$&1.44$\pm$0.45&3.45$\pm$0.69&0.84$\pm$0.22&0.87$\pm$0.22&2.21$\pm$0.42&1.28$\pm$0.20\\
\tableline
\end{tabular}
\end{center}
\tablenotetext{a}{\scriptsize Measured wavelength in \AA}
\tablenotetext{b}{\scriptsize Equivalent widths in \AA}
\tablenotetext{c}{\scriptsize Full-width at half-maximum in \kms, quadratically corrected for instrumental broadening}
\tablenotetext{d}{\scriptsize Intrinsic line flux in units of $10^{-15}\,\,{\rm erg\,\,cm}^{-2}\,\,{\rm s}^{-1}$}
\end{table*}

We reduced the XRT data with {\tt HEASOFT}~v.~6.16 and the 2014
October 2 calibration data base (CALDB) version. We used
\texttt{xrtpipeline}~v.~0.13.1 to collect events in Windowed Timing
(WT) mode to avoid pile-up. The source and background spectra were
extracted with \texttt{xselect}~v.~2.4c using 40-pixel square boxes in
the range 0.4--10~keV. We generated the ancillary response file (ARF)
with \texttt{xrtmkarf} and used the latest version (v.~015) of the
response matrices provided by the \sw\ team. We rebinned
the spectrum to obtain a minimum of 100 counts per channel. 
\newline

The UVOT photometry was obtained in all filters, {\it uvw2}, {\it
  uvm2}, {\it uvw1}, {\it u}, {\it b}, and {\it v}, and we produced an
image in each of them with {\tt uvotimsum}. We then used
{\tt uvotsource} to extract the source in a 5\arcsec\ region and the
background counts in a 15\arcsec\ source-free circular aperture,
respectively. The derived \swj\ flux densities are listed in
\autoref{uvot}.

\subsection{ATCA observations}

The array was in the compact H75 configuration and the
  observation was conducted with the Compact Array Broadband Backend
  \citep[CABB,][]{2011Wilson} and each frequency band was composed of
  2048 1-MHz channels.  We used PKS~B1934–638 for absolute flux and
  bandpass calibration, and PKS~J1741-038 to calibrate the antenna
  gains as a function of time. Flagging, calibration and imaging were
  carried out with the Multi-channel Image Reconstruction, Image
  Analysis and Display software \citep[MIRIAD,][]{1995Sault}. Due to
  poor weather conditions, most of the observation was flagged
  and \swj\ was not detected.  We obtained the following 3-$\sigma$ upper
  limits: 0.15 mJy at 5.5 GHz and 0.20 mJy at 9GHz, respectively.

\subsection{{\it WISE} observations}

For comparison only, we also make use of Wide-Field
  Infrared Survey Explorer \citep[\textit{WISE},][]{2010Wright}
  data. The source was observed with {\it WISE} at five different
  epochs, around MJD~55274, MJD~55454, MJD~55456, MJD~56736, and
  MJD~56918 (see~\autoref{wiseflux} for the weight-averaged flux
    densities). In particular, the MJD~56918 observation occurred
    about one month after our X-shooter observations (MJD~56885) and
    consists of 13 W1 (3.35~\mic) and W2 (4.5~\mic) frames of 7.7~s
    each obtained during the NEOWISE reactivation survey
    \citep{2014Mainzer} between MJD~56918.2 and MJD~56919.1
    (\autoref{wise}). These magnitudes cannot therefore be considered
    as contemporaneous to our data and this is the reason why we only
    superimpose the resulting weight-averaged flux densities on the
    modeled SEDs.

\begin{figure}
\begin{center}
\includegraphics[width=9cm]{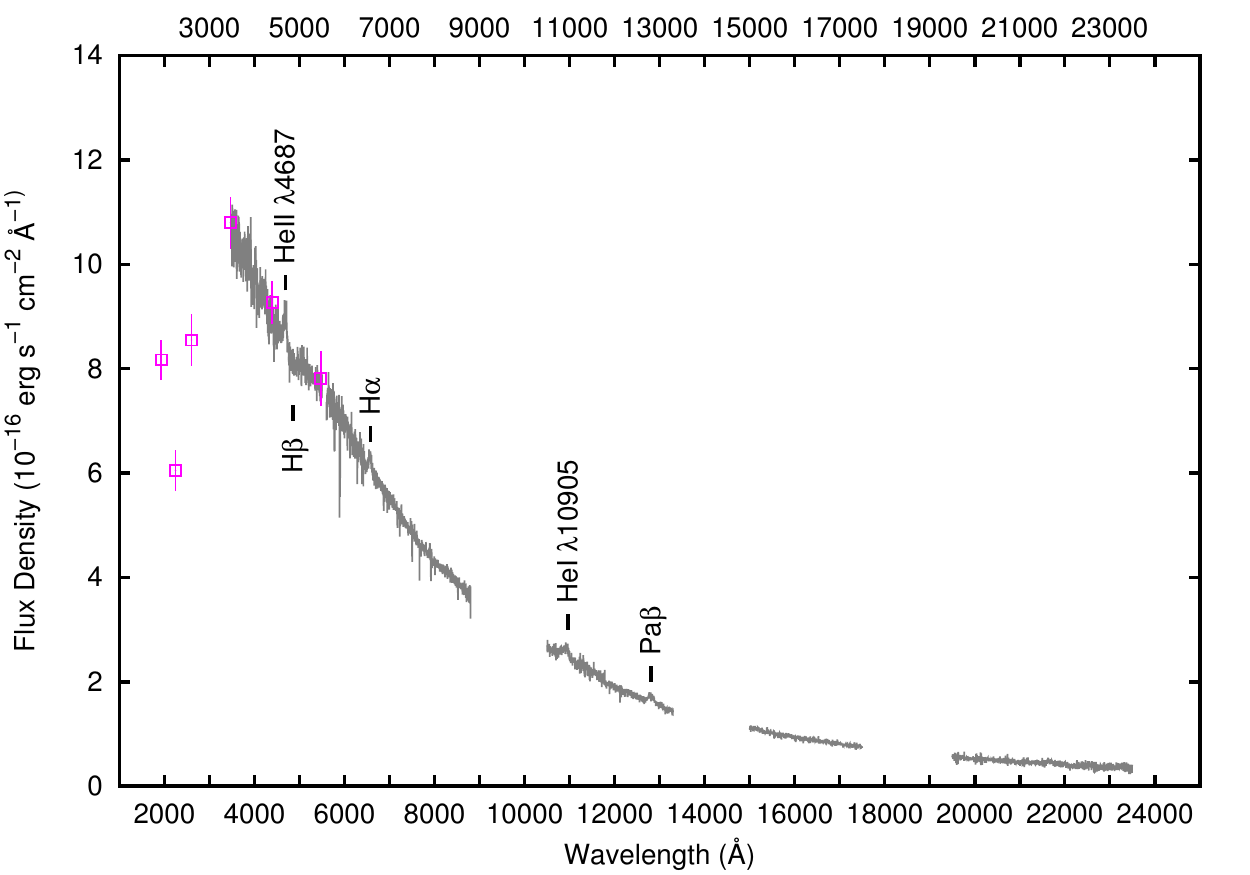}
\caption{\small Flux-calibrated  X-shooter spectrum
  of \swj, not corrected for the ISM extinction along
  the line-of-sight of the source. The detected spectroscopic lines are marked
  and the quasi-simultaneous UVOT flux densities superimposed (red).} 
\label{contspec}
\end{center}
\end{figure}

\begin{figure*}
\begin{center}
\begin{tabular}{cc}
\includegraphics[width=8cm]{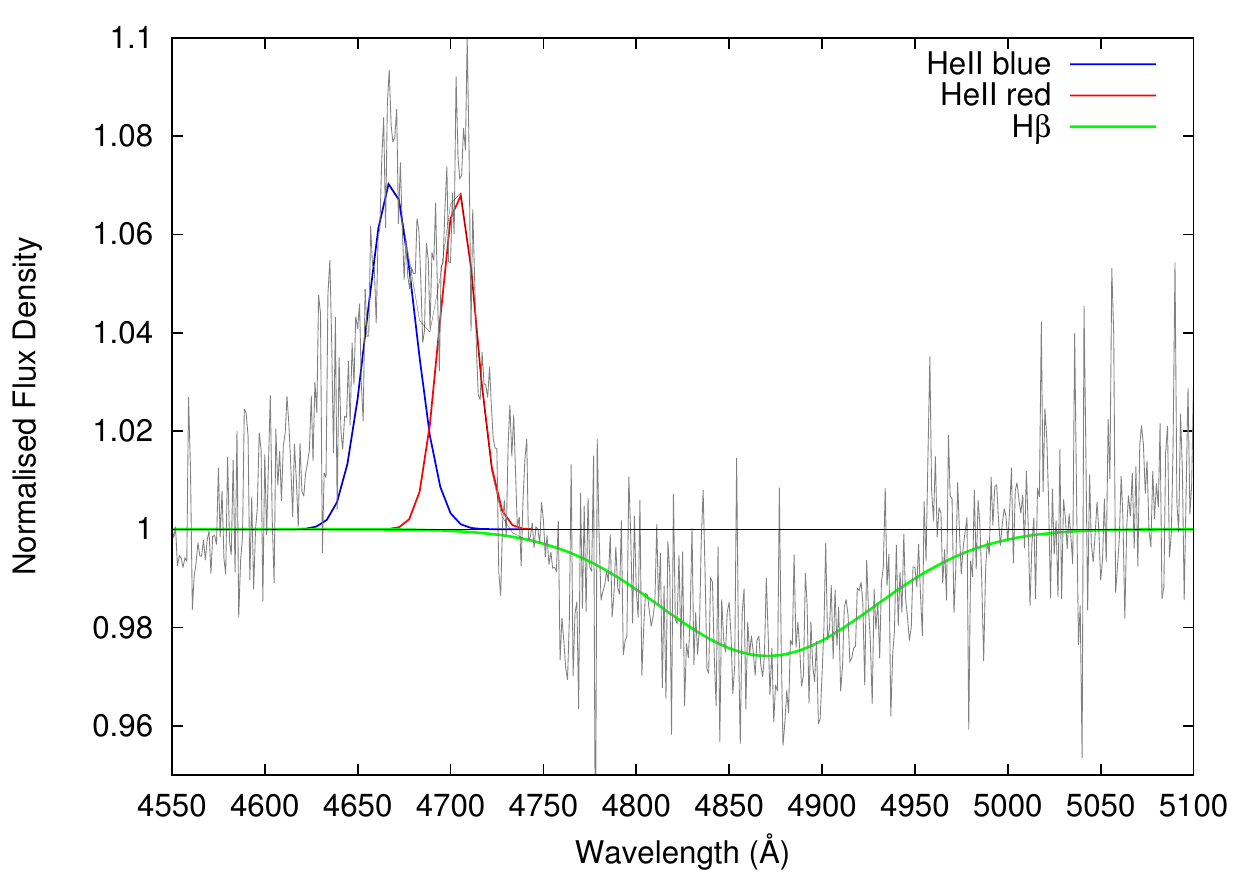}&\includegraphics[width=8cm]{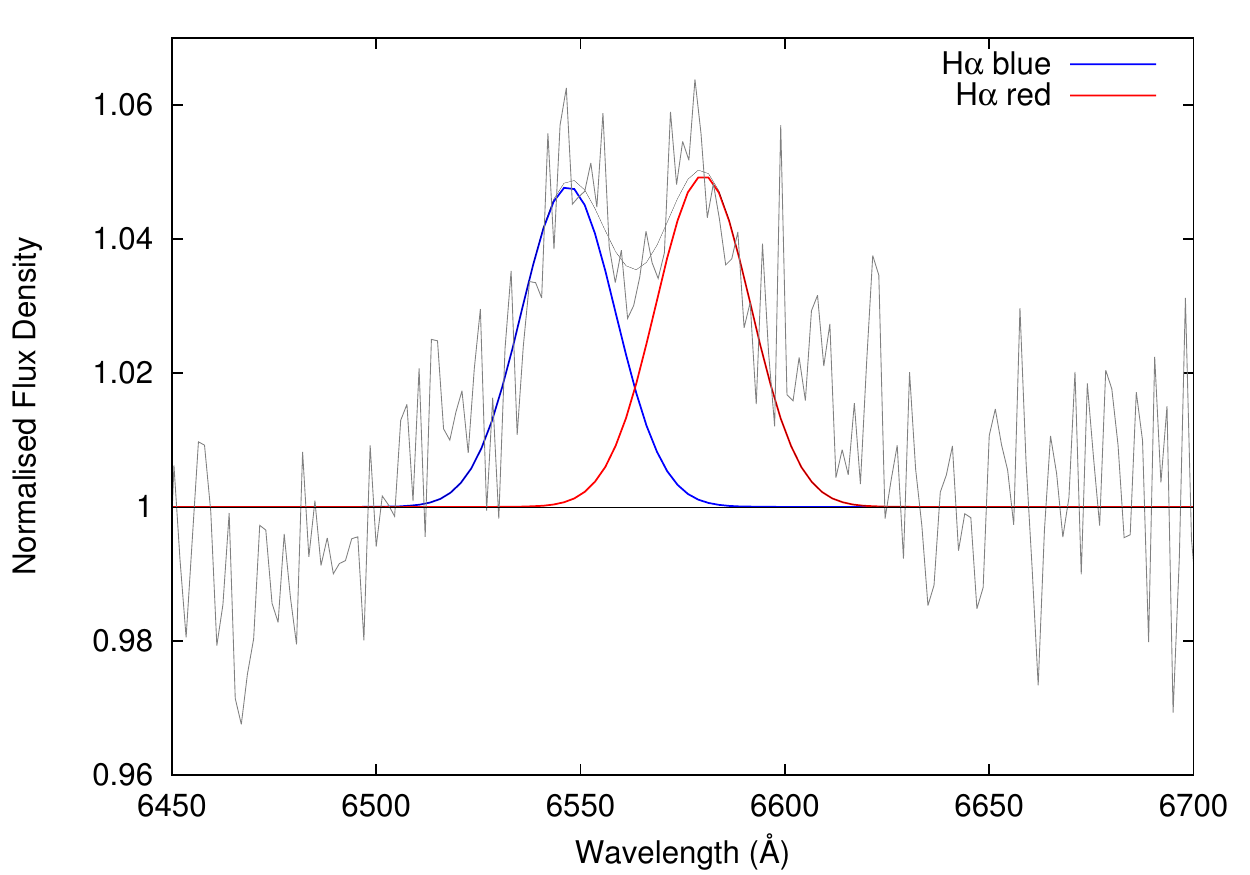}\\
\includegraphics[width=8cm]{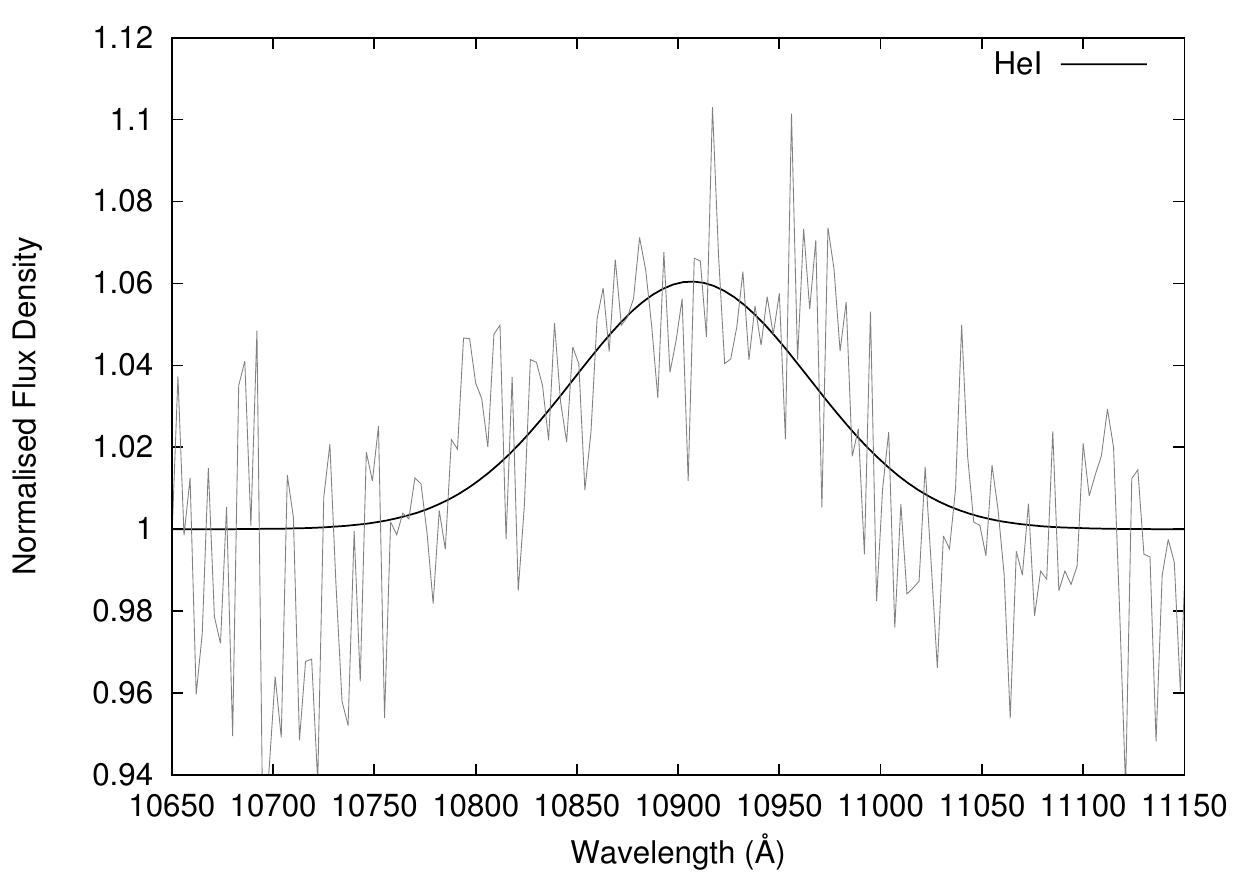}&\includegraphics[width=8cm]{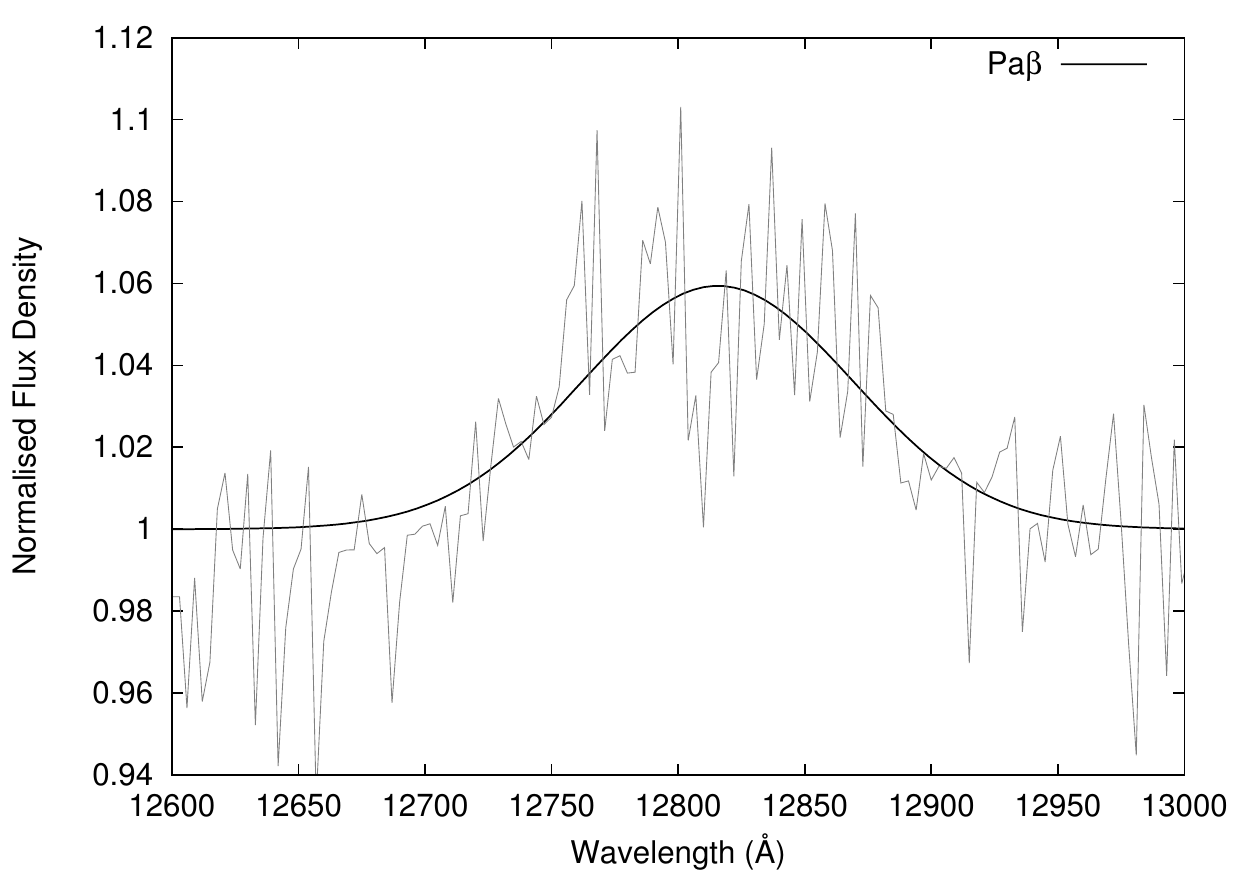}\\
\end{tabular}
\caption{\small  Optical and near-IR spectroscopic lines detected in
  the X-shooter spectrum of \swj. (Top-left): double-peaked
  \ion{He}{2}~$\lambda4686$ emission and broad H{\small $\beta$}
  absorption; (Top-right): double-peaked H{\small $\alpha$}
  emission; (Bottom-left): Broad \ion{He}{1}~$\lambda10905$ emission,
  likely double-peaked; (Bottom-right): Broad Pa{\small $\beta$}
  emission, likely double-peaked.} 
\label{irlines}
\end{center}
\end{figure*}

\section{The optical and near-IR spectrum}

 \autoref{contspec} displays the flux-calibrated X-shooter 350-2400~nm
 spectrum, on which the detected spectroscopic features are marked and
 the UVOT flux densities superimposed. \autoref{irlines} highlights
 the lines $-$ fitted with one or two Gaussian(s) $-$ and
 \autoref{speclines} lists their main parameters.  The full-widths at
 half-maximum (FWHMs) were quadratically corrected for the
 instrumental broadening and the underlying continuum was locally
 assessed with a first-order polynomial. The continuum level being the
 primary source of inaccuracy, each measurement was repeated several
 times with different continuum placements within the same wavelength
 range to obtain a set of values that eventually averaged out. The listed
 uncertainties are therefore the scatter to the mean rather than just
 statistical.

\subsection{The spectroscopic content}

In the optical domain, \citet{2014Neustroev} detected
  \ion{He}{2}~$\lambda4686$~\AA\ and \ha\ in August 2013 
  spectra. This was the first report of optical
  emission features since the very early outburst of \swj\
  \citep{2005Torres}, as both \citet{2007Cadolle}
  and \citet{2009Durant} reported the absence of optical emission lines
  during observations that took place in August 2005 and June 2007,
  respectively. \citet{2014Froning} also reported the detection of UV
    emission lines in October 2012 {\it HST} spectra. This likely
  hints at a renewed activity of the accretion disk anywhere
  between 2007 and 2012,  although we cannot be more
  accurate. \citet{2014Neustroev} showed that the two lines were 
double-peaked and the authors measured total FWHMs of 4250~\kms\ and
2450~\kms, peak-to-peak separations of 2690~\kms\ and 1650~\kms, and
total equivalent widths of 4.3~\AA\ and 3.6~\AA\ for \ion{He}{2} and
\ha, respectively. In our spectrum, we also report the two
emission lines, which are still double-peaked, and we fitted their
profiles with two Gaussians. Our measurements are in agreement with
the values given in \citet{2014Neustroev}, and we derive peak-to-peak
separations of $2304\pm286$~\kms\ and $1509\pm129$~\kms\ for
\ion{He}{2} and \ha, respectively. We also report a very broad
trough longwards of \ion{He}{2} (FWHM$\sim8800$~\kms) that is present
in all the individual spectra and that we tentatively associate with
\hb. Such Balmer absorption lines have been detected in other
microquasars \citep[e.g.][]{1995Callanan,1997Bianchini,
  2000Soria,2001Dubus,2014Rahouib} and are thought to originate in the
viscous accretion disk \citep{1989Ladous}. 
In the near-IR, we report for the first time the presence of very
broad emission lines of \ion{He}{1} and \pb\ centered at 10906~\AA\
and 12815~\AA, respectively. We believe their profiles are
double-peaked, although double-Gaussian fits are statistically not 
required for either of them due to low S/N ratios. Their FWHMs,
measured through single-Gaussian fitting, are roughly on par with the
total FHWMs of the optical emission lines, hinting at similar locations in
the accretion disc. Besides the features discussed above, no
other emission lines are detected.

\begin{table}
\begin{center}
\caption{\scriptsize Equivalent widths $\mathring{W}$ $-$ converted into $E(B-V)$ ISM reddening values $-$ of selected ISM absorption lines present in the \swj\ spectrum. We derive an average value $E(B-V)=0.45\pm0.02$.\label{diblines}}
\begin{tabular}{c|c|c}
\tableline\tableline
&$\mathring{W}$&$E(B-V)$\\
\tableline
DIB~$\lambda5780$&0.27$\pm$0.02&0.47$\pm$0.03\tmark[a]\\
\ion{Na}{1}~D2~$\lambda5890$&0.71$\pm$0.05&0.42$\pm$0.04\tmark[b]\\
\ion{Na}{1}~D1~$\lambda5896$&0.57$\pm$0.04&0.47$\pm$0.05\tmark[b]\\
\ion{K}{1}~$\lambda7699$&0.12$\pm$0.02&0.43$\pm$0.07\tmark[c]\\
\tableline
\end{tabular}
\end{center}
\tablenotetext{a}{\scriptsize Using the $\mathring{W}$ to $E(B-V)$ conversion factor given in \citet{1994Jenniskens}}
\tablenotetext{b}{\scriptsize Using the $\mathring{W}$ to $E(B-V)$ relation given in \citet{2012Poznanski}}
\tablenotetext{c}{\scriptsize Using the $\mathring{W}$ to $E(B-V)$ relation given in \citet{1997Munari}}
\end{table}

\subsection{The interstellar reddening}

\begin{figure*}
\begin{center}
\begin{tabular}{cc}
\includegraphics[width=8cm]{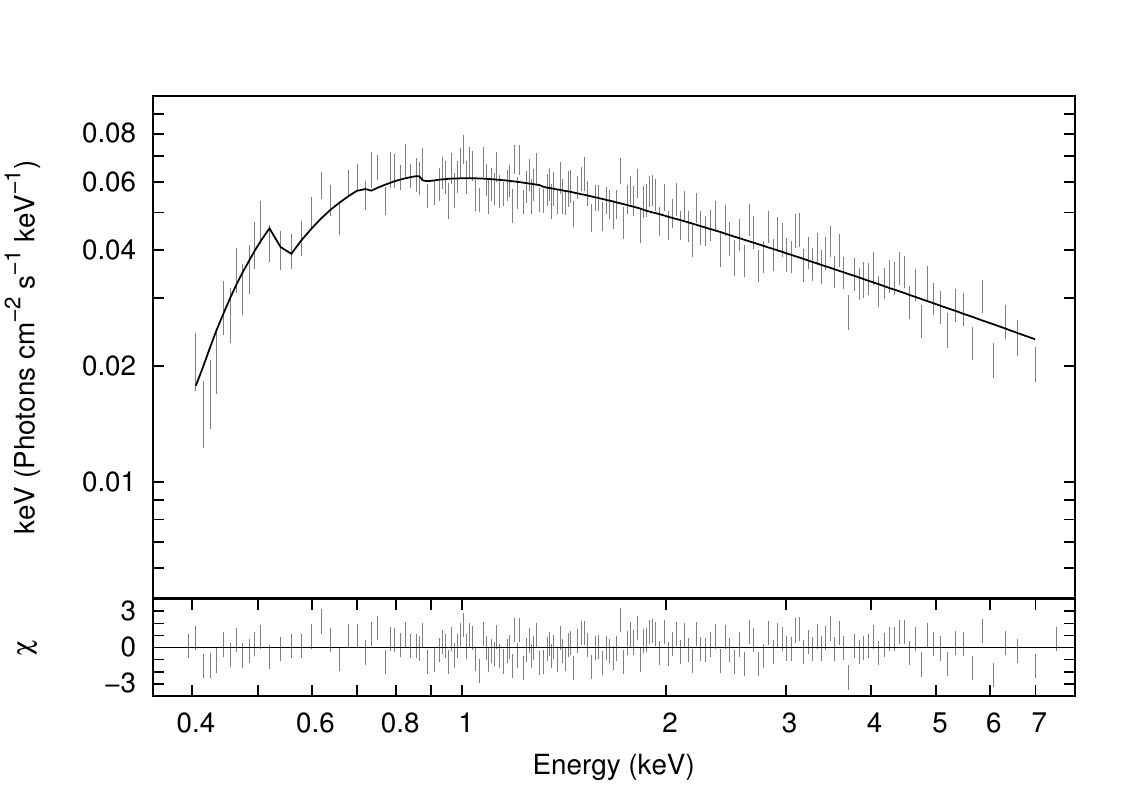}&\includegraphics[width=8cm]{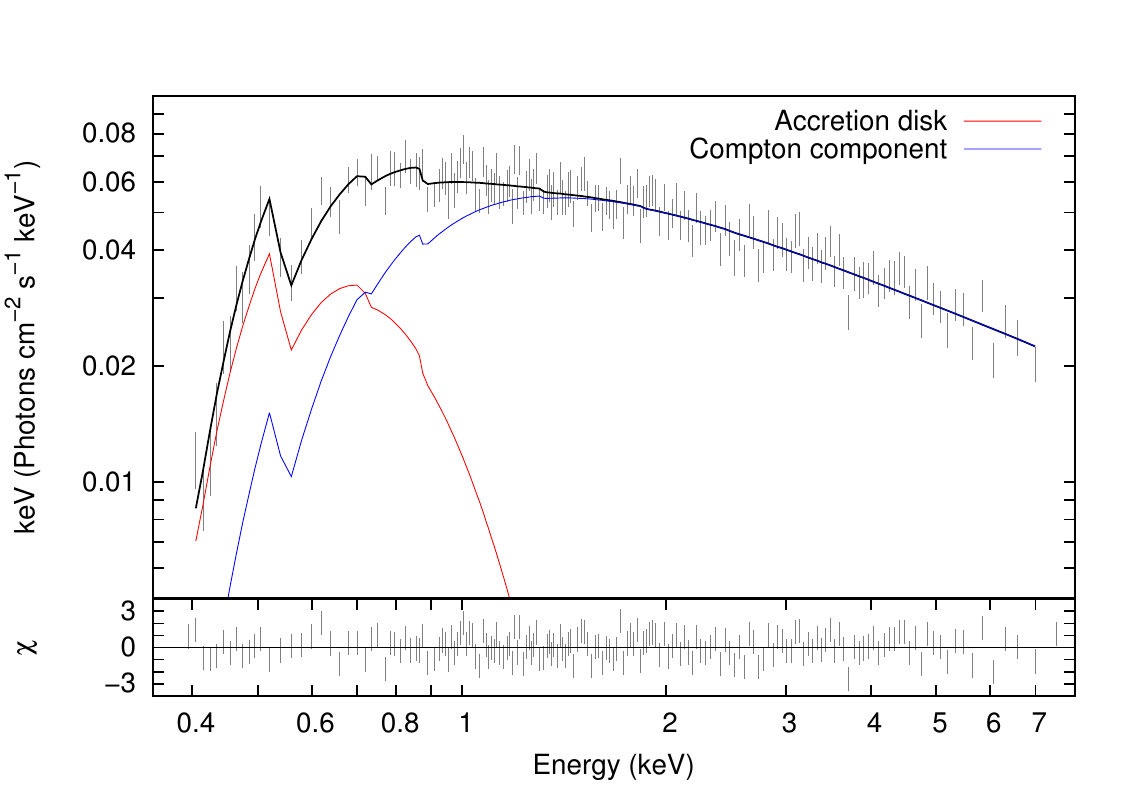}\\
\end{tabular}
\caption{\small Best-fit to the \sw/XRT spectrum of \swj\ with the models
  {\sc tbabs$\times$comptt} (left) and {\sc
    tbabs$\times$(diskbb+comptt)} (right). While the latter model only
  gives a marginally better result, the inferred
  column density is more realistic for that celestial position,
  implying a weak contribution from the accretion disk.} 
\label{bestfitx}
\end{center}
\end{figure*}

We surveyed the X-shooter spectrum to find interstellar lines
relevant for the determination of the reddening suffered by \swj\
along its line-of-sight. We unambiguously identified four of
them with S/N high enough to be used to that purpose: the diffuse
interstellar band (DIB) at 5780~\AA, the D2 and D1 \ion{Na}{1} doublet
at 5890~\AA\ and 5896~\AA, respectively, as well as the \ion{K}{1}
line at 7699~\AA. We measured their respective equivalent widths
through Gaussian fitting and used various $\mathring{W}$ vs $E(B-V)$
relations $-$ \citet{1994Jenniskens} for DIB~$\lambda5780$,
\citet{1997Munari} for \ion{K}{1}, and \cite{2012Poznanski} for 
the \ion{Na}{1} doublet $-$ to derive four $E(B-V)$ measurements that
we averaged to yield $E(B-V)=0.45\pm0.02$ (see
\autoref{diblines}). This result is similar to previous
measurements, in particular $E(B-V)=0.42\pm0.02$ found in
\citet{2009Durant} and $E(B-V)=0.45\pm0.05$ in
\citet{2014Froning}. Moreover, of relevance for multiwavelength SED
modeling, this interstellar reddening value is consistent with a
column density between $\nhe=(3.08\pm0.19)\times10^{21}$~cm$^{-2}$ and
$\nhe=(4.00\pm0.24)\times10^{21}$~cm$^{-2}$,
using the average total-to-selective extinction ratio $R_{\rm V}=3.1$
and the relations $\nhe=(2.21\pm0.09)\times10^{21}\Ave$ and
$\nhe=(2.87\pm0.12)\times10^{21}\Ave$ given in \citet{2009Guver} and
\citet{2015Foight}, respectively.

\subsection{The continuum}

\swj\ exhibits a very blue spectrum and the equivalent widths of the
near-infrared emission lines are significantly larger than those
  at optical wavelengths despite similar or lower intrinsic
fluxes. Considering that these features are likely formed in the outer
regions of the accretion disk, this points towards the disk as the
main contributor to the continuum.

Furthermore, the optical flux level is similar to those reported in
previous studies, which is consistent with the very weak average
photometric variability of the source. However, \citet{2014Neustroev}
reported a flattening of the continuum between 4000~\AA\ and 5400~\AA,
which is not present in our spectrum. The authors also argue that the
optical continuum was variable on a night-to-night timescale,
strongly between 4000~\AA\ and 5400~\AA, and marginally beyond. To
check if we could detect any similar variations, we extracted all
the individual 210~s X-shooter spectra and compared them; we could not 
find any significant flickering, neither in flux level nor in shape,
beyond the noise level. It is not clear if this can be
interpreted as a non variability of the continuum at short timescales,
or if 210~s is sufficient to smooth out the variations, like in the
\gx\ case \citep{2012Rahoui}. 

\section{Spectral energy distribution modeling}

\begin{figure*}
\begin{center}
\begin{tabular}{cc}
\includegraphics[width=8cm]{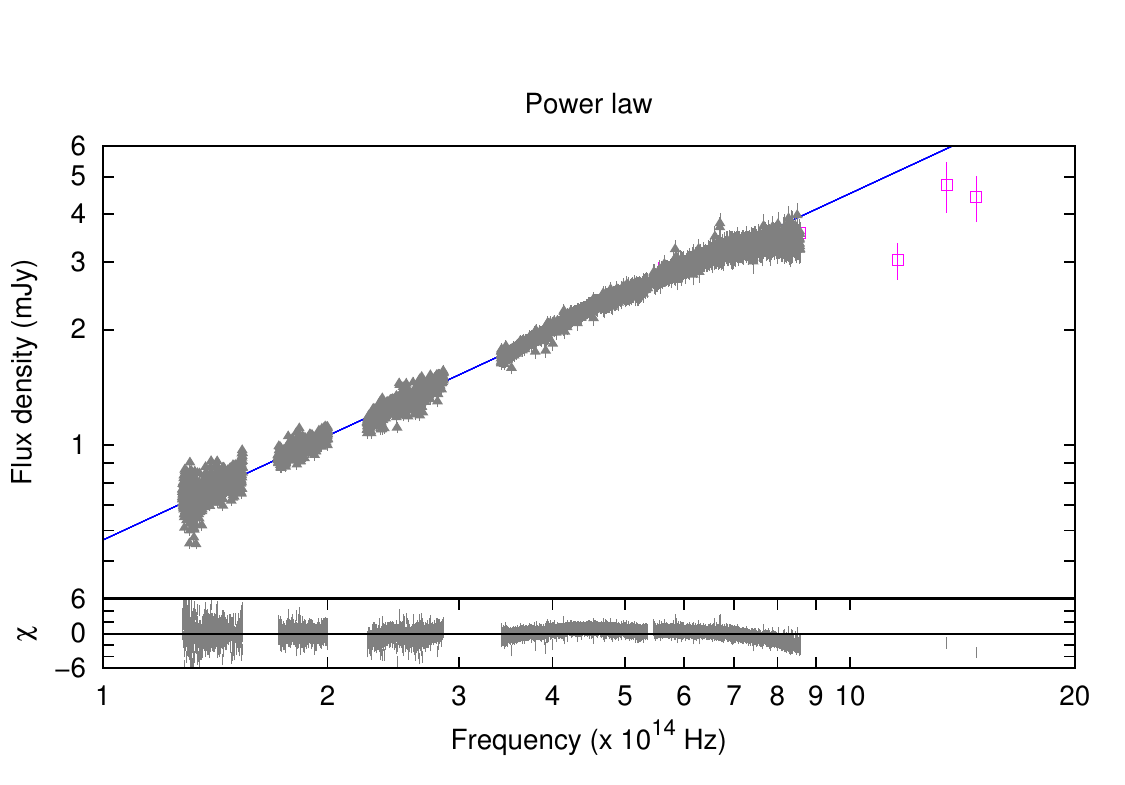}&\includegraphics[width=8cm]{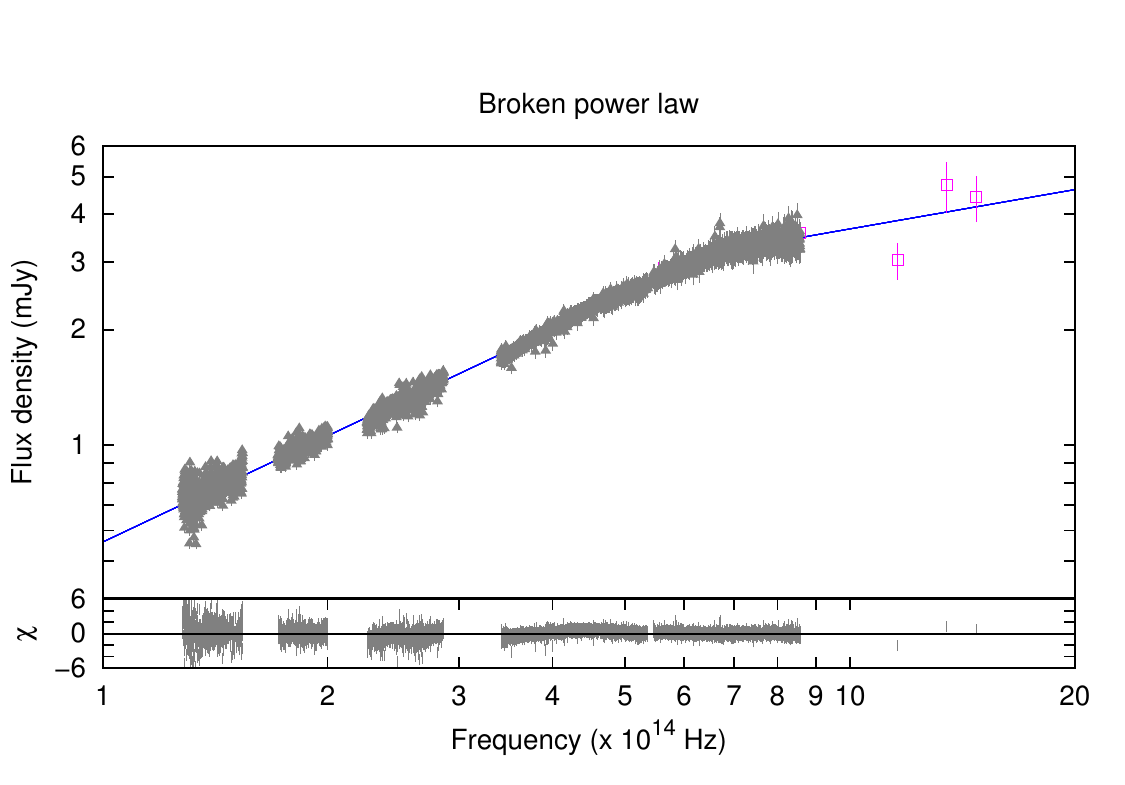}\\
\end{tabular}
\caption{\small Phenomelogical fit to the UVOT+X-shooter data using a
power law (left) and a broken power law (right). The best fit is
obtained with the broken power law and it is consistent with the
expected contribution from a viscous accretion disk, although an extra
component must be present to account for the flattening in the near-IR.} 
\label{bestfitxshooter}
\end{center}
\end{figure*}

\begin{table}
\begin{center}
\caption{\scriptsize  Best parameters obtained from fits to the
   \swj\ \sw/XRT spectrum with the two models {\sc tbabs$\times$comptt} and {\sc tbabs$\times$(diskbb+comptt}). The errorbars are given 
   at the 90\% confidence level.\label{parbestfitx}}
\begin{tabular}{ccc}
\tableline\tableline
\nh\tmark[a]&$0.14_{-0.03}^{+0.02}$&$0.35_{-0.09}^{+0.08}$\\
$kT_{\rm in}$\tmark[b]&\nodata&0.13$\pm$0.01\\
$R_{\rm in}$\tmark[c]&\nodata&$21.7_{-9.8}^{+14.1}$\\
$kT_{\rm 0}$\tmark[d]&0.12$\pm$0.02&tied to $kT_{\rm in}$\\
$kT_{\rm e}$\tmark[e]&60 (fixed)&60 (fixed)\\
$\tau$\tmark[f]&$0.91_{-0.07}^{+0.08}$&$0.76_{-0.08}^{+0.09}$\\
$\chi_{\rm r}^2$ (d.o.f)&0.91 (173)&0.86 (172)\\
\tableline
\end{tabular}
\end{center}
\tablenotetext{a}{\scriptsize Column density in units of $10^{22}$~cm$^{-2}$}
\tablenotetext{b}{\scriptsize Accretion disc inner temperature (keV)}
\tablenotetext{c}{\scriptsize Inner radius in units of $\frac{D_{\rm BH}}{M_{\rm BH}\sqrt{\cos{i}}}$~$R_{\rm g}$,
with $i$ the inclination, $D_{\rm BH}$ the distance in kpc,
$M_{\rm BH}$ the mass in \msun, and $R_{\rm g}$ the gravitational radius}
\tablenotetext{d}{\scriptsize Input soft photon temperature (keV)}
\tablenotetext{e}{\scriptsize Electron temperature (keV)}
\tablenotetext{f}{\scriptsize Electron optical depth}
\end{table}

In the following, the X-shooter spectrum and UVOT photometric points
were corrected for interstellar reddening using the extinction law
given in \citet{1999Fitzpatrick} with $E(B-V)=0.45\pm0.02$ (see
Section~3.2) and a selective-to-total extinction ratio equal to the
Galactic average value $R_{\rm V}=3.1$. 

\subsection{The X-ray emission}

We fitted the spectrum using {\tt Xspec}~v.~12.8.2, first with an
absorbed spherical Comptonization component, {\sc
  tbabs$\times$comptt} \citep{1994Titarchuk}. We used 
the abundances given in \citet{2000Wilms}, and fixed the electron
temperature to 60~keV after confirming that this parameter was not
constrained (see the T15 for a measurement of $kT_{\rm
  e}$ from our 2014 April {\it NuSTAR} observations). The best-fit
parameters are listed in the left column of \autoref{parbestfitx},
and the best-fit model is displayed in the left panel of
\autoref{bestfitx}. While the result is satisfactory, with a
seed-photons temperature $kT_{\rm 0}\sim0.12$, an optical depth 
$\tau\sim0.91$, and a reduced \chis\ value of 0.91, the measured
best-fit column density$\nhe=1.4_{-0.3}^{+0.2}\times10^{21}$~cm$^{-2}$
is a factor two lower than the expected extinction along the
line-of-sight to \swj\ (see Section 4.2 for a discussion about the
interstellar reddening). This likely means that the soft X-ray flux is
underestimated, and we thus performed the fit again adding
an accretion disk component, modeled with {\sc diskbb}. We tied the
seed-photons temperature $kT_{\rm 0}$ to the disk temperature, and
left the electron temperature fixed to 60~keV. The best-fit parameters
are listed in the right column of \autoref{parbestfitx}, and the
best-fit spectrum is displayed in the right panel of
\autoref{bestfitx}. Adding an accretion disk only slightly improves
the fit, with a reduced \chis\ of 0.86. However, the measured column
density, $\nhe=3.5_{-0.9}^{+0.8}\times10^{21}$~cm$^{-2}$ is much more 
consistent with the expected interstellar reddening. The seed-photons
blackbody included in {\sc comptt} is also negligible compared to the
disk blackbody emission. We conclude that this implies the presence of
an extra soft component contributing to the X-ray emission of \swj,
although we stress that our \sw\ data alone are not sufficient to
prove the accretion disk hypothesis.

\begin{table}
\begin{center}
\caption{\scriptsize  Best parameters obtained from fits to the
   \swj\ X-shooter spectrum with a power law and a broken power law. The errorbars are given
   at the 90\% confidence level.\label{parbestfitxshooter}}
\begin{tabular}{ccc}
\tableline\tableline
$\alpha_{\rm 1}\tmark[a]$&$0.902\pm0.002$&$0.918\pm0.002$\\
$\nu_{\rm b}$\tmark[b]&\nodata&$6.74_{-0.11}^{+0.10}\times10^{14}$\\
$\alpha_{\rm 2}$\tmark[c]&\nodata&$0.25\pm0.03$\\
$\chi_{\rm r}^2$ (d.o.f)&1.17 (6170)&0.97 (6168)\\
\tableline
\end{tabular}
\end{center}
\tablenotetext{a}{\scriptsize Spectral index of the power law below the break, if any}
\tablenotetext{b}{\scriptsize Spectral break location, in Hz}
\tablenotetext{c}{\scriptsize Spectral index of the power law beyond the break}
\end{table}

\subsection{The optical and near-IR emission}

\begin{figure}[ht!]
\begin{center}
\includegraphics[width=8cm]{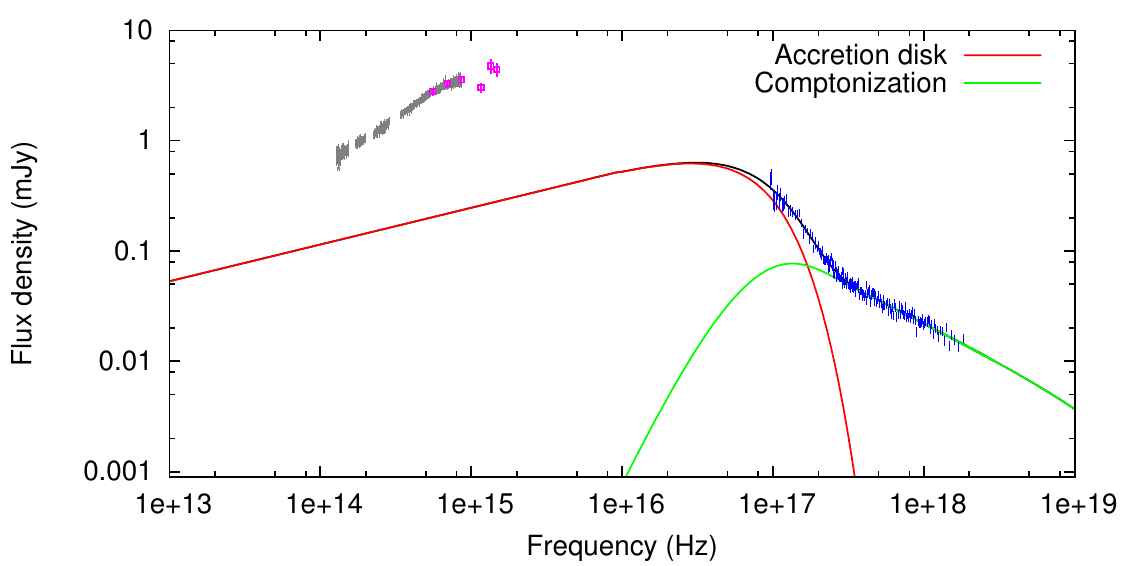}
\caption{\small Fit to the \sw/XRT+UVOT and X-shooter data using the
  X-ray-only model presented in Section~4.1. The disk emission clearly
  underestimates the low-energy flux level, which justifies the
  simultaneous fitting of the X-ray to near-IR SED to properly derive
  the accretion disk parameters. \label{bestfitxext}} 
\end{center}
\end{figure}

\citet{2014Froning} showed that the UV, optical, and near-IR emission 
of \swj\ during their 2012 October 2 observations was likely due to a
strongly truncated accretion disk, with the presence of a possible
near-IR excess. The authors also argued that the simultaneous \sw\
spectrum was perfectly modeled with an absorbed power law without any
need for a disk component. In our case, the presence of double-peaked
emission lines in the very blue X-shooter spectrum as well as the lack
of short timescale variability of the continuum also favor the
accretion disk as the main contributor to the optical and near-IR
emission. This is confirmed by the phenomelogical modeling of the
  UVOT+X-shooter continuum with a broken power law (displayed in
  \autoref{bestfitxshooter}, see \autoref{parbestfitxshooter} for the
  best-fit parameters). Beyond the break, the spectral index is
  roughly consistent with the canonical value, 1/3, expected from a
  viscous accretion disk \citep{1973Shakura}. Below, the spectral
  index is flatter than a typical Raleigh-Jeans tail, pointing towards
  the contribution of at least one extra component. 

In contrast to \citet{2014Froning}, our derived 0.3-8~keV flux of
    \swj\ is $4.35\times10^{-10}$\ergcms, i.e. about 80\% larger than
    their reported value, and we argue that an extra soft X-ray
    component is required. It is therefore likely that the disk is
    both responsible for this soft X-ray excess and a significant
    fraction of the optical and near-IR emission. Nonetheless, as seen
    in \autoref{bestfitxext}, the accretion disk continuum as derived
    from the X-ray fit only is unable to account for the UVOT and
    X-shooter flux level, and a complete characterization of the
    accretion disk properties requires a simultaneous modeling of the
    \sw/XRT+UVOT and X-shooter data.

\subsection{Broadband SED}

We therefore fitted the radio, UV/optical/near-IR, and X-ray SED of \swj\
with several models that include an accretion disk. We chose {\sc
  diskir}, introduced in \citet{2008Gierlinski}, for the following
reasons: (1) it is designed to self-consistently fit optical/near-IR
and X-ray data as it includes a Comptonization component; (2) it takes
X-ray irradiation phenomena into account; and (3) it allows a
derivation of the outer radius of the accretion disk.We also fixed a few
parameters of the {\sc diskir} model after several attempts failed to
constrain them. As previously mentioned, the electron temperature of
the Comptonization component was fixed to 60~keV. The irradiation
radius $R_{\rm irr}$, which represents the size of the inner
disk region which is illuminated by the corona, was fixed to to
  1.0001~$R_{\rm in}$, the minimum allowed value towards which it was
  converging. The fraction of hard X-ray emission thermalized in the
inner disk, $f_{\rm in}$, was frozen to the recommended value for the
hard state, 0.1, while the fraction of X-ray emission thermalized in
the outer disk, $f_{\rm out}$, which systematically converged to 0,
was eventually fixed to that value.

\begin{figure}
\begin{center}
\includegraphics[width=9cm]{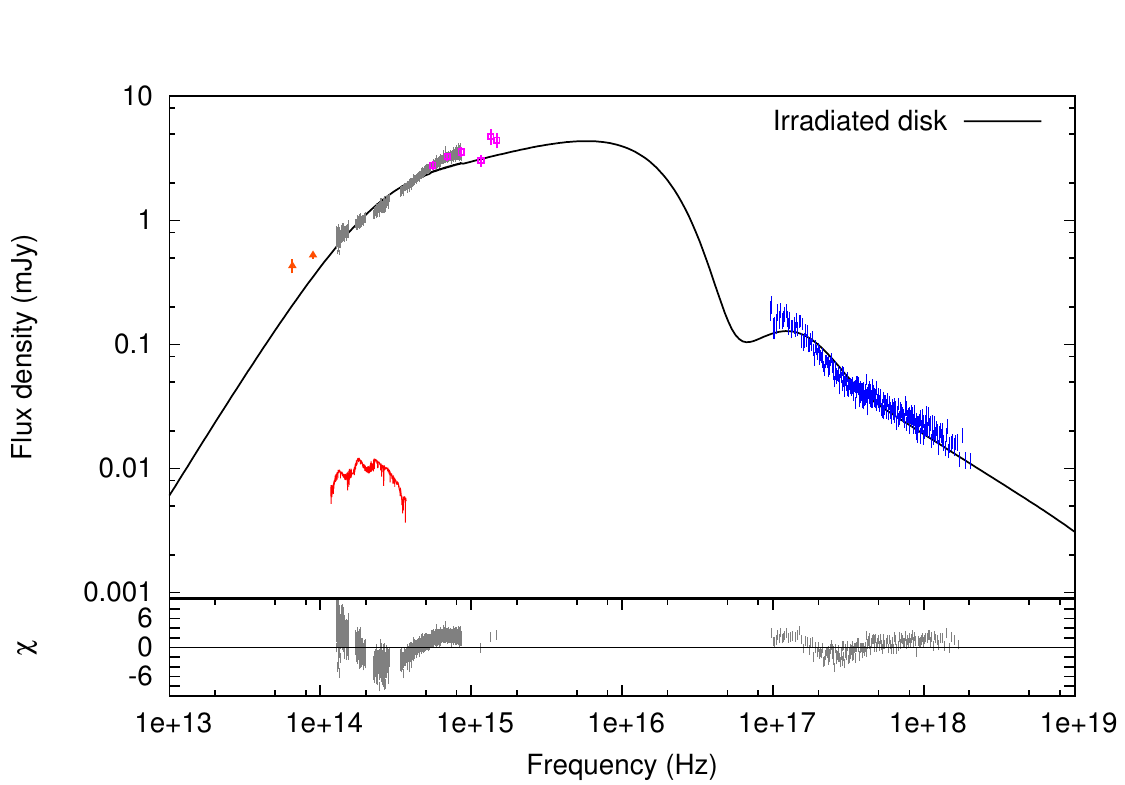}
\caption{\small Extinction-corrected VLT/X-shooter+\sw/XRT-UVOT SED of 
  \swj\ fitted with {\sc diskir} (disk-only case, see
  \autoref{parbestfit}, second column). The {\it WISE} flux densities
obtained one month after our observations (orange) are
superimposed but were not part of the fit. We also show the expected
contribution of an M4V star at a distance of 3 kpc (magenta).}
\label{disconly}
\end{center}
\end{figure}

\subsubsection{{\sc Diskir only}}

We first attempted to fit the X-shooter and XRT-UVOT SED of \swj\ with
{\sc diskir} only. \autoref{disconly} displays the best-fit model and
\autoref{parbestfit} lists the best-fit parameters. It is clear that
{\sc diskir} alone is insufficient to describe both the
X-ray and  optical/near-IR data, with a reduced \chis\
of 6.1. \autoref{disconly} also shows the superimposed
near-IR spectrum of a M4V star at  3~kpc, illustrating that no
contribution is expected from the companion star and we can therefore
confirm that besides the disk, another non-stellar
component must contribute to the X-ray and optical/near-IR
emission. We also stress that the fit gives a column density \nh\
consistent with the  interstellar reddening and points towards a cold
and truncated accretion disk.

\subsubsection{{\sc Diskir} and blackbody}

\begin{figure}
\begin{center}
\includegraphics[width=9cm]{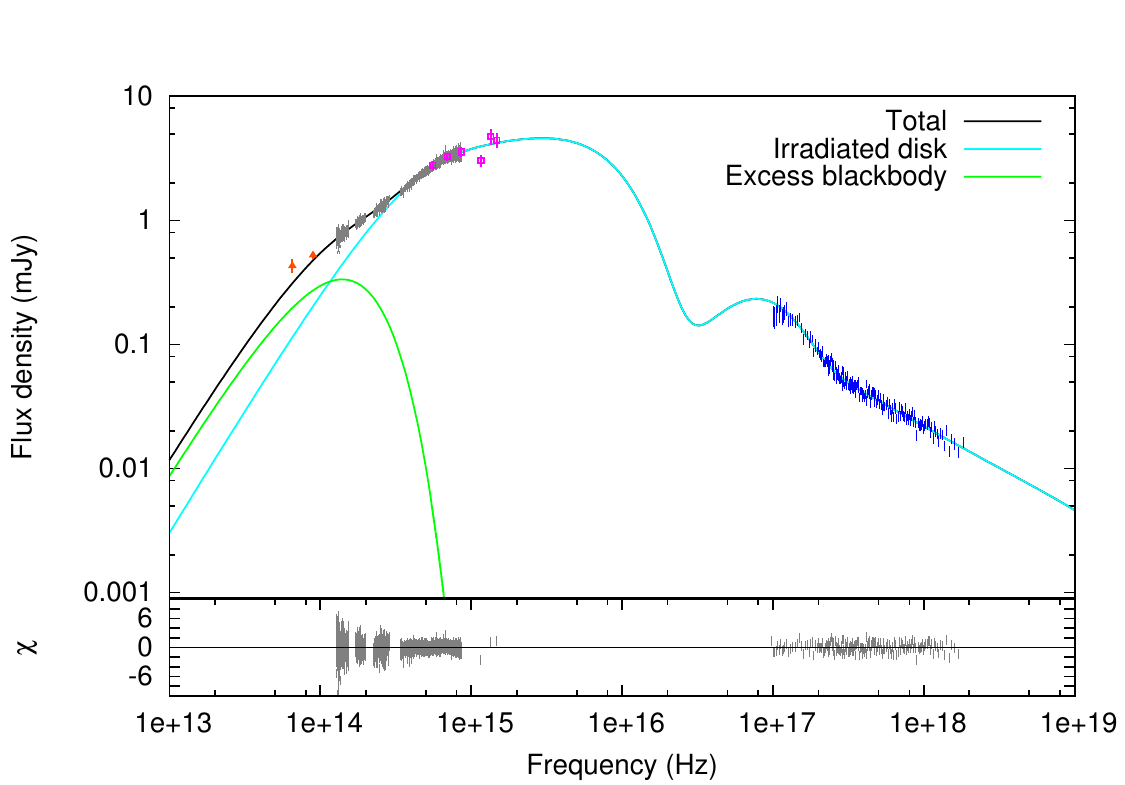}
\caption{\small Extinction-corrected VLT/X-shooter+\sw/XRT-UVOT SED of 
  \swj\ fitted with {\sc diskir+bbodyrad} (disk plus blackbody case, see
  \autoref{parbestfit}, second column). The {\it WISE} flux densities
obtained one month after our observations (orange) are
superimposed but were not part of the fit.}
\label{discbbody}
\end{center}
\end{figure}

The phenomenological modeling displayed in \autoref{bestfitxshooter}
as well as the {\sc diskir}-only fit are consistent with the presence of an
excess in the optical and near-IR domains. Although it is clear that the
emission from the companion star is negligible, the
stellar hemisphere facing the black hole could still be irradiated by the
X-ray emission, leading to such an excess. Alternatively, a warm dust
component similar to those previously detected in other microquasars
\citep[see e.g.][]{2010Rahoui} could also be present. To test these
possibilities, we fit the X-shooter and XRT-UVOT SED of \swj\ adding
a spherical blackbody to {\sc diskir}; the best-fit parameters are
listed in the second column of \autoref{parbestfit} and the best-fit
SED is displayed in \autoref{discbbody}. The addition of the blackbody
emission clearly improves the fit, with a reduced \chis\ of 0.94,
confirming the presence of an excess in the optical and near-IR
domains. However, the best-fit parameters are neither consistent with
stellar irradiation nor warm dust. Indeed, the best-fit temperature,
$2376\pm25$~K, is too high to be that of warm dust, which sublimes
around 1500~K \citep[see e.g.][]{2003Draine}, while the best-fit
radius, $2.72\pm0.03$~\rsun\ for a distance of 3~kpc, is too large to be
that of an M dwarf star, typically smaller that 0.6~\rsun\
\citep{2012Boyajian}. As such, the optical and near-IR excess must
therefore stem from another process.

\subsubsection{{\sc Diskir} and broken power law}

\begin{figure}[!ht]
\begin{center}
\includegraphics[width=9cm]{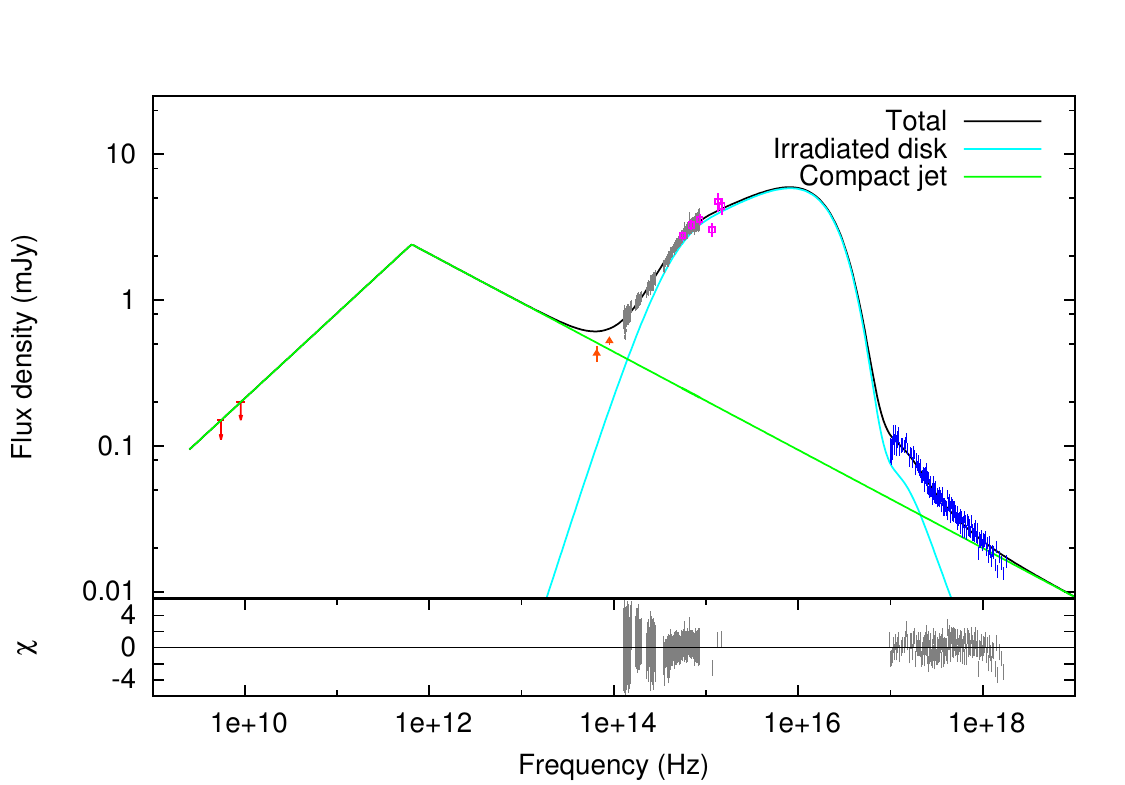}
\caption{\small Extinction-corrected ATCA+VLT/X-shooter+\sw/XRT-UVOT
  SED of \swj\ fitted with {\sc diskir+bknpow} (Case~1., see
  \autoref{parbestfit}). Here we fixed the optically
  thick compact jet spectral index to 0.59, i.e. the value derived for
  the two ATCA upper limits at 5.5 and 9 GHz. The {\it WISE} flux densities
obtained one month after our observations (orange) are
superimposed but were not part of the fit.}
\label{bestfit_po}
\end{center}
\end{figure}

\begin{figure}[!ht]
\begin{center}
\includegraphics[width=9cm]{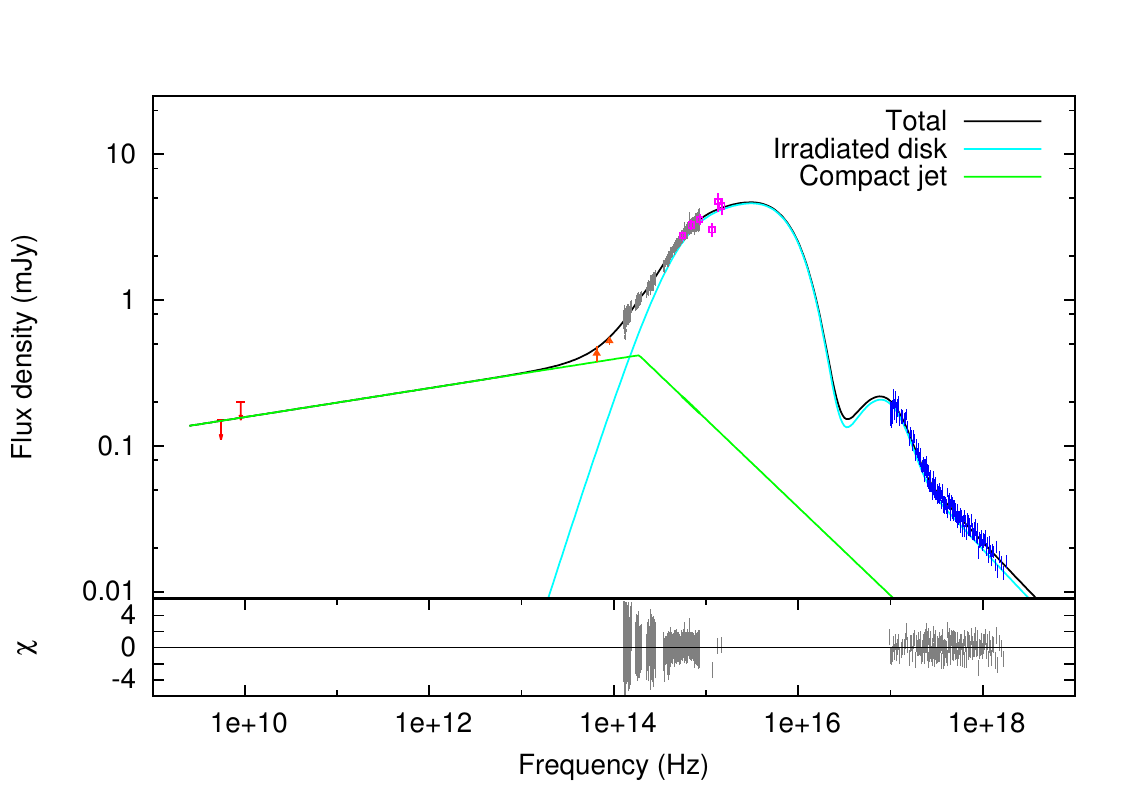}
\caption{\small Extinction-corrected ATCA+VLT/X-shooter+\sw/XRT-UVOT SED of
  \swj\ fitted with {\sc diskir+bknpow} (Case~2., see
  \autoref{parbestfit}). Here the optically thick compact jet
spectral index is not fixed and the fit lets it to be flat. The {\it WISE} flux densities
obtained one month after our observations (orange) are
superimposed but were not part of the fit.}
\label{bestfit}
\end{center}
\end{figure}

Our observations occurred when \swj\ was in the hard state and the
source is regularly detected at radio frequencies. In particular, its
radio flux at 15.4 GHz was measured with AMI at about 290 $\mu$Jy a few
days after our observations (M. Kolehmainen, private
communication).  It is therefore reasonable to expect
that a compact jet contributes to its optical and near-IR emission through
synchrotron radiation, as seen in many other microquasars \citep[see
e.g.][]{1998Eikenberry, 2002Corbel, 2003Chaty, 2006Russell,
  2010Russell}. To test this hypothesis, we replaced the
  blackbody by a broken power law to mimic the bolometric emission
  of the jet as calculated in \citet{1979Blandford}, i.e. a
  combination of optically thick synchrotron
  ($F_{\nu}\propto\nu^{\alpha_{\rm 1}\ge0}$) from the radio domain to
  a spectral break $\nu_{\rm b}$ where the synchrotron becomes optically
  thin ($F_{\nu}\propto\nu^{\alpha_{\rm 2}\le0}$). Unfortunately,
  \swj\ was not detected at radio frequencies during our observations
  due to poor conditions, and we can only rely on radio upper limits to
  constrain the optically thick spectral index $\alpha_{\rm
    1}$. Here, we therefore consider the two
  following cases: (1) we fix $\alpha_{\rm 1}=0.59$, i.e. the
    value derived from the radio upper limits at 5.5 and 9~GHz; and
  (2) $\alpha_{\rm 1}$ is left free to vary, but the broken
  power law is forced to match at least the upper limit at 5.5~GHz; the 
  derived break frequencies must consequently be considered as lower
  limits. The best-fit models for the two cases are
  displayed in \autoref{bestfit_po} and \autoref{bestfit},
  respectively, and \autoref{parbestfit} lists the best-fit parameters. 

  Case~2 clearly is the best fit, with a reduced \chis\ of 
  0.94, and is also the most phenomenologically relevant. Indeed, the
  derived parameters point towards a weak contribution to the XRT
  spectrum from a cold and truncated disk, while the coronal emission
  strongly dominates the X-ray emission. Likewise, the optical and
  near-IR spectrum is well described by the thermal emission from the
  viscous accretion disk, with an excess due to the synchrotron
  emission from the jet, both optically thick and thin, as the spectral
  break is located around $1.88\times10^{14}$~Hz or 1.60 \mic. We
  nonetheless stress here that the location of the spectral break in the
  near-IR domain may be an artificial effect of the fitting process
  due to the lack of information between the radio and X-shooter
  data. We refer the reader to the next section for a discussion on
  the reliability of this location and the good agreement between the
  fit and the {\it WISE} flux densities obtained about a month after our
    observations.

  In contrast to Case~2, Case~1 is worse,
  with a reduced \chis\ of 1.11. More importantly, it is
  also less relevant from a physical point of view as the derived
  ratio between the Comptonization component and the unilluminated
  disk is inconsistent with hard state spectra of microquasars, with
  $L_{\rm c}/L_{\rm d}\approx0.11$, whereas a value larger than 1 is
  expected. A possible explanation is that the fitting process tends
  to overestimate the contribution from the jet at the expense of that
  from the corona. Furthermore, while theoretically not ruled out
    \citep{1979Blandford}, an optically thick synchrotron spectral
    index of 0.6 has, to our knowledge, never been observed for
    \swj. In \citet{2011Soleri}, the authors report on several radio
    spectral indices observed between 2005 and 2009 and find a maximum
    index of about 0.3, which is similar to the value we report in
    T15. 

\begin{table*}
\caption{\scriptsize Best-fit parameters derived from fitting of \swj\ X-shooter+\sw/XRT-UVOT SED with {\sc tbabs$\times$diskir} (left) and {\sc tbabs$\times$(diskir+bknpow)} (right). Error bars are given at the 90\% confidence level.\label{parbestfit}}
\begin{center}
\begin{tabular}{cc|c|cc}
\tableline\tableline
&{\sc tbabs$\times$diskir}&{\sc tbabs$\times$(diskir+bbody)}&\multicolumn{2}{c}{{\sc tbabs$\times$(diskir+bknpow)}}\\
\tableline
&&&Case~1&Case~2\\
\nh\tmark[a]&$0.32\pm$0.01&$0.30\pm0.01$&$0.21\pm0.02$&$0.29\pm0.02$\\
$kT_{\rm disc}$\tmark[b]&$0.027_{-0.001}^{+0.002}$&$0.014\pm0.001$&$0.037\pm0.008$&$0.015_{-0.001}^{+0.002}$\\
$\Gamma$\tmark[c]&$1.78\pm0.02$&$1.70\pm0.03$&$2.71_{-0.26}^{+0.30}$&$1.70\pm0.03$\\
$kT_{\rm e}$\tmark[d]&$60$ (fixed)&60 (fixed)&$60$ (fixed)&$60$ (fixed)\\
$L_{\rm c}/L_{\rm d}$\tmark[e]&$1.42_{-0.17}^{+0.14}$&$3.79_{-0.44}^{+0.62}$&$0.11_{-0.02}^{+0.07}$&$3.22_{-0.67}^{+0.64}$\\
$f_{\rm in}$\tmark[f]&0.10 (fixed)&0.10 (fixed)&0.10 (fixed)&0.10 (fixed)\\
$f_{\rm out}$\tmark[g]&0 (fixed)&0 (fixed)&0 (fixed)&0 (fixed)\\
$R_{\rm in}$\tmark[h]&$(6.03\pm0.46)\times10^{2}$&$1.71_{-0.15}^{+0.28}\times10^{3}$&$4.47_{-0.88}^{+2.88}\times10^{2}$&$1.60_{-0.19}^{+0.26}\times10^{3}$\\
$R_{\rm out}$\tmark[i]&$345.1_{-23.7}^{+30.3}$&$63.8_{-8.7}^{+6.7}$&$223.8_{-66.6}^{+63.9}$&$57.3_{-7.6}^{+8.3}$\\
$R_{\rm irr}$\tmark[j]&1.0001 (fixed)&1.0001 (fixed)&1.0001 (fixed)&1.0001 (fixed)\\
$\alpha_{\rm 1}\tmark[k]$&\nodata&\nodata&$0.59$ (fixed)&$0.10\pm0.01$\\
$\nu_{\rm b}$\tmark[l]&\nodata&\nodata&$6.41_{-0.32}^{+0.35}\times10^{11}$&$1.88_{-0.04}^{+0.02}\times10^{14}$\\
$\alpha_{\rm 2}$\tmark[m]&\nodata&\nodata&$-0.34\pm0.01$&$-0.60\pm0.06$\\
$T_{\rm BB}$\tmark[n]&\nodata&2376$\pm$25&\nodata&\nodata\\
$R_{\rm BB}$\tmark[o]&\nodata&0.91$\pm0.01$&\nodata&\nodata\\
$\chi_{\rm r}^2$ (d.o.f)&6.09 (6344)&0.93 (6342)&1.11 (6344)&0.94 (6343)\\
\tableline
\end{tabular}
\end{center}
\tablenotetext{a}{\scriptsize Column density in units of $10^{22}$~cm$^{-2}$}
\tablenotetext{b}{\scriptsize Temperature of the unilluminated accretion disk in keV}
\tablenotetext{c}{\scriptsize Asymptotic power law photon index}
\tablenotetext{d}{\scriptsize Electron temperature in keV}
\tablenotetext{e}{\scriptsize Ratio of luminosity of the Compton component with respect to the unilluminated disk}
\tablenotetext{f}{\scriptsize Fraction of luminosity in the Compton tail which is thermalized in the inner disk}
\tablenotetext{g}{\scriptsize Fraction of the total luminosity thermalized in the outer disk}
\tablenotetext{h}{\scriptsize Inner radius in units of $\frac{D_{\rm BH}}{M_{\rm BH}\sqrt{\cos{i}}}$~$R_{\rm g}$, with $i$ the inclination, $D_{\rm BH}$ the distance in kpc, $M_{\rm BH}$ the mass in \msun, and $R_{\rm g}$ the gravitational radius}
\tablenotetext{i}{\scriptsize Outer radius in units of $R_{\rm in}$}
\tablenotetext{j}{\scriptsize Radius of the Compton illuminated disk in units of $R_{\rm in}$}
\tablenotetext{k}{\scriptsize Spectral index of the optically thick synchrotron emission of the compact jet}
\tablenotetext{l}{\scriptsize Spectral break of the compact jet in Hz}
\tablenotetext{m}{\scriptsize Spectral index of the optically thin synchrotron emission of the compact jet}
\tablenotetext{n}{\scriptsize Blackbody excess temperature in K}
\tablenotetext{o}{\scriptsize Blackbody excess radius in \rsun/kpc}
\end{table*}

\section{Discussion}

The following discussion is based on our best-fit results for Case~2
and, unless stated otherwise, we assume the following
system parameters: (1) 40\degr\ inclination; (2) a $q=0.04$ mass
ratio \citep[lower limits,][]{2014Neustroev}; (3) $M_{\rm
  {BH}}=5$~\msun\ BH mass; (4) $a=1.53$~\rsun\ semi-major
axis \citep[upper limits,][]{2014Neustroev}; and (5) 3~kpc distance,
which roughly corresponds to the expected value for a 5~\msun\ BH
\citep{2014Froning}. Using the approximate formulas given in
\citet{1983Eggleton} and \citet{2002Frank}, we derive a Roche lobe
radius $R_{\rm {L}}\sim0.64a=6.81\times 10^{10}$~cm, a tidal
radius $R_{\rm {tide}}\sim0.58a=6.18\times 10^{10}$~cm,
and a circularization radius $R_{\rm {circ}}\sim0.46a=4.90\times
10^{10}$~cm. Finally, in the Keplerian approximation, the velocity of
the accretion disk region that most contributes to a given Gaussian
line is related to its
FWHM in \kms\ as
\begin{equation}
V=\frac{FWHM}{2\times\sqrt{\ln{2}}\times\sin{i}}\,\,\rm{km}\,\rm{s}^{-1}
\label{vrot}
\end{equation}
If the line is double-peaked, we also can derive the Keplerian
velocity from the peak-to-peak separation $\Delta{v}$ following
\begin{equation}
V=\frac{\Delta{v}}{2\times\sin{i}}\,\,\rm{km}\,\rm{s}^{-1}
\label{vrot2}
\end{equation}
The Keplerian radius of the region that most contributes to the lines
is then given by
\begin{equation}
R=\frac{c^2}{V^2}\,\,R_{\rm  g}
\label{rrot}
\end{equation}
where $R_{\rm  g}\,=\,GM_{\rm BH}/c^2$ is the gravitational radius. 

\subsection{The accretion disk properties}

All our SED fits listed in \autoref{parbestfit} point towards a very
truncated accretion disk. From our best fit, we infer an inner radius
$R_{\rm in}\sim1097$~$R_{\rm g}\,=\,8.2\times10^8$~cm and an outer
radius $R_{\rm out}\sim6.3\times10^4\,\,R_{\rm
  g}=4.66\times10^{10}$~cm for case~2. If we assume that the trough
detected in the optical spectrum is H{\small $\beta$} absorption from
the disk, then \autoref{vrot} and \autoref{rrot} also lead to $R_{\rm
  in}\le1343\,\,R_{\rm g}$. Likewise, assuming H{\small
  $\alpha$} originates from the outer accretion disk, we 
  infer $R_{\rm out}\sim6.6\times10^4\,\,R_{\rm
  g}=4.89\times10^{10}$~cm from both its FWHM and peak-to-peak
separation. Both our SED modeling and X-shooter spectrum are therefore
consistent with each other and point towards a small and very
truncated accretion disk, with $R_{\rm in}\sim10^3$~$R_{\rm g}$. Such
a large truncation is possible for an ADAF-like flow, but it is
thought to be typical of the quiescent state of microquasars, during
which their luminosity  drops below $10^{-5}\,L_{\rm edd}$
\citep{2008Narayan}. In contrast, the \swj\ bolometric luminosity
during our observation is about $0.005\,L_{\rm  edd}$, 
more typical of a relatively faint hard state. Interestingly, another
outlier, GRO J0422+32, was found to have an even larger 
  disk/ADAF transition radius \citep{1998Esin} ; this
  might hint at the presence of very large truncations in the
hard state of outliers.

The low number of emission lines, likely all double-peaked,
in the spectrum of \swj\ is in stark contrast to the wealth of
emission lines, mostly single-peaked, detected in the hard state
optical/near-IR spectra of other microquasars \citep[see
e.g.][]{1997Bandy, 2001Wu, 2014Rahouib}. Outer regions of
accretion disks can be directly illuminated by the hard X-ray
emission from the Comptonization component and/or the compact jets in
the hard  state, resulting into strong UV/optical/near-IR excess.
The paucity of emission lines as well as the fact that a significant
fraction of the X-shooter continuum can be explained solely by the
thermal emission from the viscous accretion disk therefore hints at a
very low level of outer accretion disk irradiation. 

Possible reasons are the low X-ray luminosity or the outer accretion
disk not being flared up. However, even a weak X-ray illumination of
the accretion disk chromosphere is thought to create an inflated envelope
\citep{1983Begelman} that, in the hard state, would 
likely play an important role in reflecting a significant fraction of
the X-ray emission back to the outer regions, where it would be
thermalized \citep{1983Begelmanb, 2002Jimenez,
  2009Gierlinski}. Following \citet{1983Begelman}, such an envelope
could exist in a quasi-hydrostatic equilibrium at any radius $R$
in the disk if $T_{\rm c}<T_{\rm g}$, where 
\begin{equation}
T_{\rm c}\,=\,\frac{1}{4k_{\rm B}L_{\rm X}}\int_{\nu_1}^{\nu_{\rm n}} h\nu L_{\nu}
  d\nu\,\,\textrm{K} 
\label{tc}
\end{equation}
 is the Compton temperature, with $L_{\rm
     X}=\int_{\nu_1}^{\nu_{\rm n}} L_{\nu}d\nu$ the irradiating X-ray
   luminosity, and 
\begin{equation}
T_{\rm g}\,=\,\frac{2}{3} \frac{G\,M_{\rm BH}\,m_{\rm p}}{k_{\rm
    B}\,R}\,\,\textrm{K} 
\label{tg}
\end{equation}
is the escape temperature. Equating \autoref{tc} and \autoref{tg}
leads to the typical radius $R_{\rm c}$ within which the envelope is
in quasi-hydrostatic equilibrium 
\begin{equation}
R_{\rm c}\,=\,\frac{2}{3} \frac{\,m_{\rm p}\,c^2}{k_{\rm B}\,T_{\rm
    c}}\,\,R_{\rm g} 
\label{rc}
\end{equation}
also called the Compton radius. For the \swj\ parameters and X-ray
emission, this results in $T_{\rm c}\approx2.8\times10^8$~K and
$R_{\rm   c}\approx2.6\times 10^4$~$R_{\rm g}$. In the \gx\ case, in
which a high level of X-ray irradiation is found \citep{2014Rahouib},
the same calculations lead to $T_{\rm c}\approx6.6\times10^7$~K and
$R_{\rm c}\approx1.2\times10^5$~$R_{\rm g}$ compared to $R_{\rm 
  out}\approx1.8\times10^5$~$R_{\rm g}$. The quasi-hydrostatic
envelope therefore covers about 40\% of the accretion disk in \swj\
versus 67\% in \gx. Even if other parameters
may account for the low level of thermalized hard X-ray photons,
a reasonable explanation is thus that a significant fraction
of hard X-ray photons do not reach the outer regions of the
accretion disk. It is also interesting that the Compton
  temperature is more than four times larger in \swj\ than \gx\,
  despite the latter being ten times more X-ray luminous than the
  former ($0.005\,L_{\rm edd}$ vs $0.05\,L_{\rm edd}$, respectively). In
  the X-ray spectrum of \gx\ in the hard state,
  a significant contribution from the accretion disk is present
  \citep{2012Rahoui}. In contrast, the \swj\ X-ray emission mostly
  stems from Comptonization (see the next section and T15). This
  actually illustrates the fact that Compton heating is as sensitive
  to the hardness of the X-ray spectrum as to the X-ray luminosity
  itself (see \autoref{tc}). This is why bright hard states of
  microquasars could be more favorable than soft states to Compton
  heating of accretion disk chromospheres and the presence of
  thermally driven winds launched from the envelope \citep[see][for a
  discussion on \gx]{2014Rahouib}. Nonetheless, a necessary condition
  is that the irradiating X-ray emission is at least twice as bright than a
  critical luminosity defined as 
\begin{equation}
  L_{\rm cr}\,=\,\frac{288}{\sqrt{T_{\rm c}}}L_{\rm
  edd}. 
\end{equation}
The presence of Compton-heated winds can therefore be ruled
out for \swj\ as $L_{\rm cr}\approx0.02\,L_{\rm edd}$, to compare to a
luminosity of about $0.005\,L_{\rm edd}$.

\subsection{The compact jet emission and the origin of the X-ray emission} 

\begin{figure}
\begin{center}
\includegraphics[width=9cm]{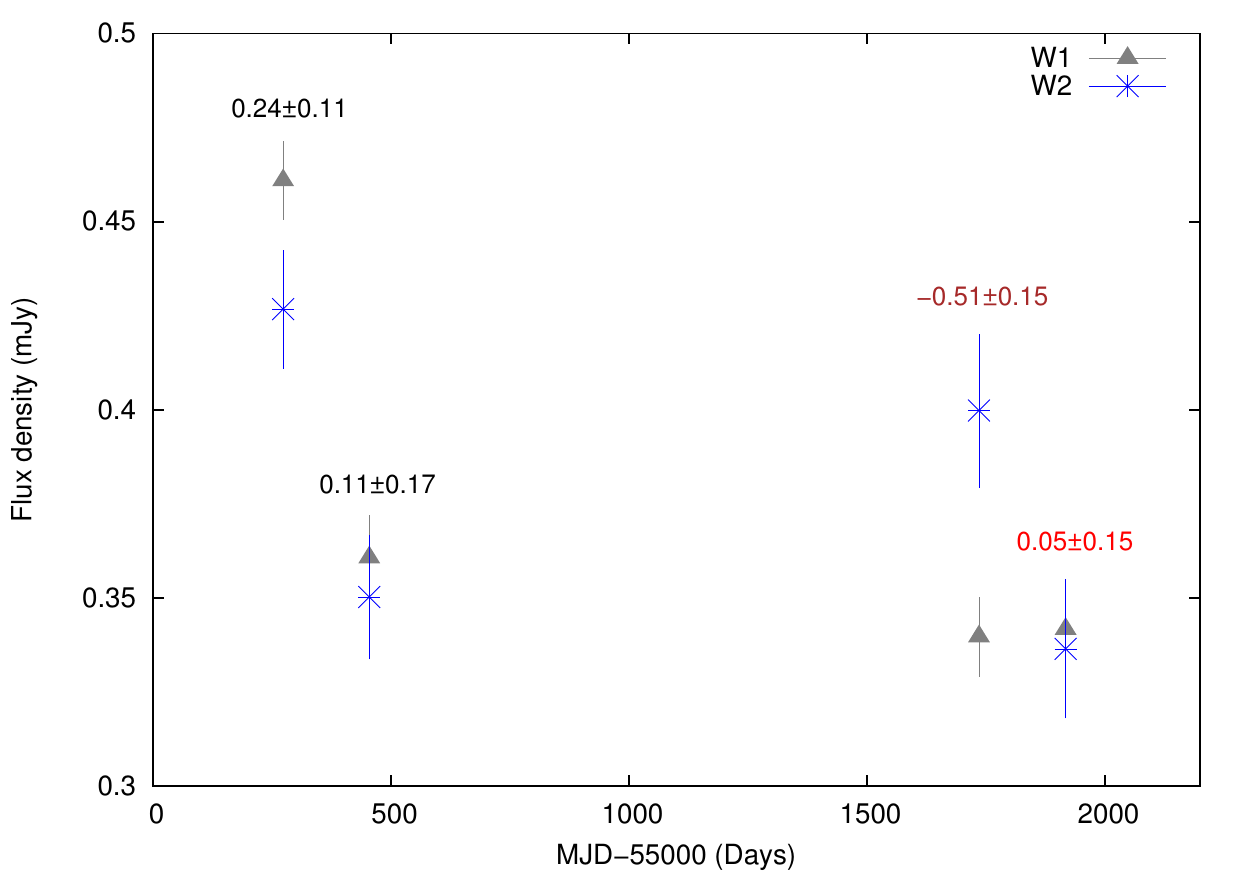}
\caption{\small Dereddened {\it WISE} W1 and W2 flux
  densities obtained on MJD~55274, MJD~55454, MJD~56736, and
  MJD~56918. They are corrected for the contribution from the
  accretion disk as measured in  Case~2 to trace the compact jet
  emission. We also give the spectral index derived at each epoch,
  including 20 days before our April observations (brown, see T15) and
  one month after the August observations reported in this paper (red).}
\label{var_wise}
\end{center}
\end{figure}

The SED modeling points towards the detection of the
compact jet in the near-IR via, at least, an optically thin
synchrotron component well-constrained by the presence of a near-IR
excess. But whether optically thick synchrotron emission
contributes to the optical/near-IR domain depends on the frequency
turnover $\nu_{\rm b}$ and the non-detection of the source in the
radio domain strongly limits our capacity to constrain its
location although it is very likely beyond $6.4\times10^{11}$~Hz.
We note that the W1 and W2 {\it WISE} flux
densities obtained about one month after our observations are almost
consistent with our Case~2 scenario, for which the spectral 
  break is located in the near-IR domain, beyond
  $1.9\times10^{14}$~Hz. If a coincidence cannot be ruled out, this
  could also mean that the average compact jet emission was relatively
  constant over a few months and that Case~2 properly describes the
  jet properties. To check to which extent such a weakly-inverted
  infrared emission is common, we measured the mid-infrared slopes at
  other epochs based on the archival {\it WISE}
  data. \autoref{var_wise} displays the dereddened W1 and W2 {\it 
    WISE} flux densities of \swj\ for MJD~55274, MJD~55454,
  MJD~56736, and MJD~56918; note that we do not include the
  measurements for MJD~55456 as they are similar to those on
  MJD~55454. These flux densities were first corrected for the
  accretion disk contribution as measured in our fits. We thus
  assume that the accretion disk emission is relatively constant,
  which is reasonable considering the low variability of the optical
  magnitudes of the source since the decay of the initial outburst
  \citep[see Figure~1 in][]{2013Shaw}. Assuming that no other
  component, in particular dust, contributed \citep[see
  e.g.][]{2010Rahoui, 2012Chaty}, these flux densities therefore trace
  the compact jet emission and we can estimate the spectral index of
  its synchrotron radiation at each epoch. Besides MJD~56918 data that 
  we already show in our fits,  we find that this index was also
  positive and consistent with optically thick synchrotron on
  MJD~55274 and MJD~55454 (hence MJD~55456) which strengthens the 
  possibility that the spectral break was indeed located in the
  near-IR domain during our observations. In contrast, the spectral
  index is negative and consistent with optically thin
  synchrotron only once, on MJD~56736. Interestingly, these {\it
    WISE} data were obtained about 20 days before our April 2014
  observations. This is consistent with the results presented in T15,
  where we find that the compact jet synchrotron emission in the {\it
    WISE} bands was optically thin, and where we report a spectral
break in the range $2.4\times10^{10}-3.6\times10^{12}$~Hz, derived
with much better radio constraints when the X-ray luminosity of \swj\
was about 50\% lower. It is therefore very likely that the turnover
shifted to higher frequencies in August compared to April, especially
considering that the near-IR excess was much larger, hinting at a more
important contribution from the compact jet. Such behavior may
  be expected as $\nu_{\rm b}\propto L_{\rm X}^{1/3}$ or $\nu_{\rm
  b}\propto L_{\rm X}^{2/3}$ whether we consider a radiatively
efficient or inefficient flow, respectively \citep{1995Falcke,
  2003Heinz}. However, \citet{2013Russell} did not find any
  correlation between the spectral break frequency and the X-ray
  luminosity, and other parameters such as the base radius of the jet
and/or the strength of the magnetic field must be considered
\citep[see e.g.][]{2011Chaty, 2011Gandhi}; a detailed analysis is out
of the scope of this paper and we refer to T15 for a comprehensive
discussion of the compact jet energetics.

Nonetheless, whichever spectral break frequency we
consider, it is clear that the optically thin synchrotron from the
compact jet marginally contributes to the soft X-ray spectrum, which
almost completely stems from the Comptonization component. This is a
property that \swj\ shares with other outliers, such as
XTE~J1650$-$500 \citep{2004Corbelb}, XTE~J1720$-$318,
\citep{2006Chatya} and  Cygnus~X-1 \citep{2011Rahoui}. Outliers are
characterized by a steeper X-ray/radio correlation $L_{\rm R} \propto
L_{\rm X}^\xi$ so that the radio flux density for a given X-ray
luminosity is systematically lower than what is expected. A possible
explanation, dubbed the radio quiet hypothesis, is that compact
jets in outliers are intrinsically weaker. On the other hand,
\citet{2011Coriat} estimated $\xi\approx1.4$ for H1743$-$322 and
proposed that the accretion flow in outliers may rather be radiatively
efficient ($L_{\rm X} \propto \dot{M}$) as opposed to ``standard''
microquasars for which it is radiatively inefficient ($L_{\rm
  X}\propto \dot{M}^{2-3}$); this is the X-ray bright
hypothesis. An alternative explanation still consistent with the X-ray
bright phenomenon is however the presence of an extra component in
the soft X-ray band that is not related to the jet but contributes to the
X-ray emission, effectively creating an X-ray excess. What this
component could be is a matter of debate, but the geometry of the
system, i.e. a highly truncated accretion disk and a large ADAF-like
Comptonization component,  could be consistent with the presence of
the condensation-induced residual inner accretion disk predicted
by evaporation models \citep{2001Liu, 2007Liu, 2014Meyerh}. These
residual disks could be irradiated by the hard X-rays but would only
carry a very small fraction of the total accreted material. They 
  could in consequence be responsible for an excess soft X-ray
emission while contributing very little to the fueling of the compact
jet. Finally, \citet{2011Coriat} showed that H1743$-$322 
  transitions back to the ``standard'' radio/X-ray track below a
critical luminosity, and this behavior was also observed in
XTE~J1550$-$564 \citep{2010Russell} and XTE~J1752$-$533
\citep{2012Ratti}. This could also be explained by the presence of a
residual inner disk, as the model predicts that such a disk can only
exist between about $(0.001-0.02)\,L_{\rm Edd}$ \citep[see][for a more
detailed discussion on the effects of residual disks on the
X-ray/radio correlation in microquasars]{2014Meyerh}.

\section{Summary and conclusion}

 We have presented a multiwavelength study of the outlier \swj\ that
focused primarily on its X-shooter spectrum. Based on the spectral
analysis of double-peaked emission lines as well as SED modeling, we
find that the optical and near-IR emission of the source mostly stems
from the thermal radiation of a very truncated accretion disk, with
$R_{\rm in}\sim1000R_{\rm g}$. There also is a significant
contribution from the compact jet, the spectral break of which may be
located in the near-IR. Nonetheless,
its optically thin synchrotron radiation cannot account for the soft
X-ray emission of \swj, which mainly originates from a very large
ADAF-like Comptonization component. Finally, the level of irradiation
of the outer accretion disk is low, and we propose that this may be
due to the relative compactness of a hard X-ray-induced envelope above
the disk plane that cannot reflect enough X-ray photons back to
the outer regions. 

Although the presence of strongly truncated accretion disks in the
hard state of microquasars is still a matter of debate, our results
and those presented in T15 are consistent with previous
multiwavelength studies of \swj\ that all hint at large inner radii
\cite[see e.g.][]{2010Zhang, 2014Froning}. On the other hand, several
authors claimed the detection of a cold accretion disk extending to
the ISCO \citep{2009Reis,
  2010Reynolds, 2013Mostafa}, based solely on X-ray data, in
particular the detection of iron emission lines. These seemingly
contradictory results can be reconciled if we consider the presence
of both a strongly truncated disk and a residual one created by
condensation of the Comptonization component. This illustrates the
importance of quasi-simultaneous X-ray and optical/near-IR datasets 
to constrain the properties of accretion disks. We therefore recommend
further multiwavelength observations of outliers to understand to
which extent the presence of residual inner disks may be a common
pattern.

\section*{Acknowledgments}

We thank the referee for her/his very insightful and constructive
comments. FR thanks the ESO staff who performed the service
observations. JAT acknowledges partial support from NASA under {\em
  Swift} Guest Observer grants NNX13AJ81G and NNX14AC56G. SC
acknowledges the financial support from the UnivEarthS Labex programme
of Sorbonne Paris Cit\'e (ANR-10-LABX-0023 and ANR-11-IDEX-0005-02),
and from the CHAOS project  ANR-12-BS05-0009 supported by the French
Research National Agency. EK acknowledges support from TUBITAK BIDEB
2219 program. This work was supported by the Spanish
Ministerio de Econom\'ia y Competitividad (MINECO) under grant
AYA2013-47447-C3-1-P (SM). This research has made use of data obtained
from the High Energy Astrophysics Science Archive Research Center
(HEASARC), provided by NASA's Goddard Space Flight Center. This
publication also makes use of data products from NEOWISE, which is a
project of the Jet Propulsion Laboratory/California Institute of
Technology, funded by the Planetary Science Division of the National
Aeronautics and Space Administration. The Australia Telescope Compact
Array is part of the Australia Telescope which is funded by the
Commonwealth of Australia for operation as a National Facility managed
by CSIRO. This research has made use of NASA's Astrophysics Data
System, of the SIMBAD, and VizieR databases operated at CDS,
Strasbourg, France.

\bibliographystyle{apj}
\bibliography{../mybib_tot}{}

\begin{thebibliography}{82}
\expandafter\ifx\csname natexlab\endcsname\relax\def\natexlab#1{#1}\fi

\bibitem[{{Bandyopadhyay} {et~al.}(1997){Bandyopadhyay}, {Shahbaz}, {Charles},
  {van Kerkwijk}, \& {Naylor}}]{1997Bandy}
{Bandyopadhyay}, R., {Shahbaz}, T., {Charles}, P.~A., {van Kerkwijk}, M.~H., \&
  {Naylor}, T. 1997, MNRAS, 285, 718

\bibitem[{{Begelman} \& {McKee}(1983)}]{1983Begelmanb}
{Begelman}, M.~C., \& {McKee}, C.~F. 1983, ApJ, 271, 89

\bibitem[{{Begelman} {et~al.}(1983){Begelman}, {McKee}, \&
  {Shields}}]{1983Begelman}
{Begelman}, M.~C., {McKee}, C.~F., \& {Shields}, G.~A. 1983, ApJ, 271, 70

\bibitem[{{Bianchini} {et~al.}(1997){Bianchini}, {della Valle}, {Masetti}, \&
  {Margoni}}]{1997Bianchini}
{Bianchini}, A., {della Valle}, M., {Masetti}, N., \& {Margoni}, R. 1997, A\&A,
  321, 477

\bibitem[{{Blandford} \& {Konigl}(1979)}]{1979Blandford}
{Blandford}, R.~D., \& {Konigl}, A. 1979, ApJ, 232, 34

\bibitem[{{Boyajian} {et~al.}(2012){Boyajian}, {von Braun}, {van Belle},
  {McAlister}, {ten Brummelaar}, {Kane}, {Muirhead}, {Jones}, {White},
  {Schaefer}, {Ciardi}, {Henry}, {L{\'o}pez-Morales}, {Ridgway}, {Gies}, {Jao},
  {Rojas-Ayala}, {Parks}, {Sturmann}, {Sturmann}, {Turner}, {Farrington},
  {Goldfinger}, \& {Berger}}]{2012Boyajian}
{Boyajian}, T.~S., {et~al.} 2012, ApJ, 757, 112

\bibitem[{{Burrows} {et~al.}(2005){Burrows}, {Hill}, {Nousek}, {Kennea},
  {Wells}, {Osborne}, {Abbey}, {Beardmore}, {Mukerjee}, {Short}, {Chincarini},
  {Campana}, {Citterio}, {Moretti}, {Pagani}, {Tagliaferri}, {Giommi},
  {Capalbi}, {Tamburelli}, {Angelini}, {Cusumano}, {Br{\"a}uninger}, {Burkert},
  \& {Hartner}}]{2005Burrows}
{Burrows}, D.~N., {et~al.} 2005, \ssr, 120, 165

\bibitem[{{Cadolle Bel} {et~al.}(2007){Cadolle Bel}, {Rib{\'o}}, {Rodriguez},
  {Chaty}, {Corbel}, {Goldwurm}, {Frontera}, {Farinelli}, {D'Avanzo}, {Tarana},
  {Ubertini}, {Laurent}, {Goldoni}, \& {Mirabel}}]{2007Cadolle}
{Cadolle Bel}, M., {et~al.} 2007, ApJ, 659, 549

\bibitem[{{Callanan} {et~al.}(1995){Callanan}, {Garcia}, {McClintock}, {Zhao},
  {Remillard}, {Bailyn}, {Orosz}, {Harmon}, \& {Paciesas}}]{1995Callanan}
{Callanan}, P.~J., {et~al.} 1995, ApJ, 441, 786

\bibitem[{{Chaty} \& {Bessolaz}(2006)}]{2006Chatya}
{Chaty}, S., \& {Bessolaz}, N. 2006, A\&A, 455, 639

\bibitem[{{Chaty} {et~al.}(2011){Chaty}, {Dubus}, \& {Raichoor}}]{2011Chaty}
{Chaty}, S., {Dubus}, G., \& {Raichoor}, A. 2011, A\&A, 529, A3+

\bibitem[{{Chaty} {et~al.}(2003){Chaty}, {Haswell}, {Malzac}, {Hynes},
  {Shrader}, \& {Cui}}]{2003Chaty}
{Chaty}, S., {Haswell}, C.~A., {Malzac}, J., {Hynes}, R.~I., {Shrader}, C.~R.,
  \& {Cui}, W. 2003, MNRAS, 346, 689

\bibitem[{{Chaty} \& {Rahoui}(2012)}]{2012Chaty}
{Chaty}, S., \& {Rahoui}, F. 2012, ApJ, 751, 150

\bibitem[{{Corbel} {et~al.}(2013){Corbel}, {Coriat}, {Brocksopp}, {Tzioumis},
  {Fender}, {Tomsick}, {Buxton}, \& {Bailyn}}]{2013Corbel}
{Corbel}, S., {Coriat}, M., {Brocksopp}, C., {Tzioumis}, A.~K., {Fender},
  R.~P., {Tomsick}, J.~A., {Buxton}, M.~M., \& {Bailyn}, C.~D. 2013, MNRAS,
  428, 2500

\bibitem[{{Corbel} \& {Fender}(2002)}]{2002Corbel}
{Corbel}, S., \& {Fender}, R.~P. 2002, ApJL, 573, L35

\bibitem[{{Corbel} {et~al.}(2004){Corbel}, {Fender}, {Tomsick}, {Tzioumis}, \&
  {Tingay}}]{2004Corbelb}
{Corbel}, S., {Fender}, R.~P., {Tomsick}, J.~A., {Tzioumis}, A.~K., \&
  {Tingay}, S. 2004, \apj, 617, 1272

\bibitem[{{Corbel} {et~al.}(2000){Corbel}, {Fender}, {Tzioumis}, {Nowak},
  {McIntyre}, {Durouchoux}, \& {Sood}}]{2000Corbel}
{Corbel}, S., {Fender}, R.~P., {Tzioumis}, A.~K., {Nowak}, M., {McIntyre}, V.,
  {Durouchoux}, P., \& {Sood}, R. 2000, A\&A, 359, 251

\bibitem[{{Coriat} {et~al.}(2011){Coriat}, {Corbel}, {Prat}, {Miller-Jones},
  {Cseh}, {Tzioumis}, {Brocksopp}, {Rodriguez}, {Fender}, \&
  {Sivakoff}}]{2011Coriat}
{Coriat}, M., {et~al.} 2011, MNRAS, 414, 677

\bibitem[{{Draine}(2003)}]{2003Draine}
{Draine}, B.~T. 2003, ARA\&A, 41, 241

\bibitem[{{Dubus} {et~al.}(2001){Dubus}, {Kim}, {Menou}, {Szkody}, \&
  {Bowen}}]{2001Dubus}
{Dubus}, G., {Kim}, R.~S.~J., {Menou}, K., {Szkody}, P., \& {Bowen}, D.~V.
  2001, ApJ, 553, 307

\bibitem[{{Durant} {et~al.}(2008){Durant}, {Gandhi}, {Shahbaz}, {Fabian},
  {Miller}, {Dhillon}, \& {Marsh}}]{2008Durant}
{Durant}, M., {Gandhi}, P., {Shahbaz}, T., {Fabian}, A.~P., {Miller}, J.,
  {Dhillon}, V.~S., \& {Marsh}, T.~R. 2008, ApJL, 682, L45

\bibitem[{{Durant} {et~al.}(2009){Durant}, {Gandhi}, {Shahbaz}, {Peralta}, \&
  {Dhillon}}]{2009Durant}
{Durant}, M., {Gandhi}, P., {Shahbaz}, T., {Peralta}, H.~H., \& {Dhillon},
  V.~S. 2009, MNRAS, 392, 309

\bibitem[{{Eggleton}(1983)}]{1983Eggleton}
{Eggleton}, P.~P. 1983, ApJ, 268, 368

\bibitem[{{Eikenberry} {et~al.}(1998){Eikenberry}, {Matthews}, {Murphy},
  {Nelson}, {Morgan}, {Remillard}, \& {Muno}}]{1998Eikenberry}
{Eikenberry}, S.~S., {Matthews}, K., {Murphy}, Jr., T.~W., {Nelson}, R.~W.,
  {Morgan}, E.~H., {Remillard}, R.~A., \& {Muno}, M. 1998, ApJL, 506, L31

\bibitem[{{Esin} {et~al.}(1998){Esin}, {Narayan}, {Cui}, {Grove}, \&
  {Zhang}}]{1998Esin}
{Esin}, A.~A., {Narayan}, R., {Cui}, W., {Grove}, J.~E., \& {Zhang}, S.-N.
  1998, \apj, 505, 854

\bibitem[{{Falcke} \& {Biermann}(1995)}]{1995Falcke}
{Falcke}, H., \& {Biermann}, P.~L. 1995, A\&A, 293, 665

\bibitem[{{Fender} {et~al.}(2004){Fender}, {Belloni}, \& {Gallo}}]{2004Fenderb}
{Fender}, R.~P., {Belloni}, T.~M., \& {Gallo}, E. 2004, MNRAS, 355, 1105

\bibitem[{{Fitzpatrick}(1999)}]{1999Fitzpatrick}
{Fitzpatrick}, E.~L. 1999, PASP, 111, 63

\bibitem[{{Foight} {et~al.}(2015){Foight}, {Guver}, {Ozel}, \&
  {Slane}}]{2015Foight}
{Foight}, D., {Guver}, T., {Ozel}, F., \& {Slane}, P. 2015, ArXiv e-prints
  1504.07274

\bibitem[{{Frank} {et~al.}(2002){Frank}, {King}, \& {Raine}}]{2002Frank}
{Frank}, J., {King}, A., \& {Raine}, D.~J. 2002, {Accretion Power in
  Astrophysics} (Cambridge University Press)

\bibitem[{{Freudling} {et~al.}(2013){Freudling}, {Romaniello}, {Bramich},
  {Ballester}, {Forchi}, {Garc{\'{\i}}a-Dabl{\'o}}, {Moehler}, \&
  {Neeser}}]{2013Freudling}
{Freudling}, W., {Romaniello}, M., {Bramich}, D.~M., {Ballester}, P., {Forchi},
  V., {Garc{\'{\i}}a-Dabl{\'o}}, C.~E., {Moehler}, S., \& {Neeser}, M.~J. 2013,
  A\&A, 559, A96

\bibitem[{{Froning} {et~al.}(2014){Froning}, {Maccarone}, {France}, {Winter},
  {Robinson}, {Hynes}, \& {Lewis}}]{2014Froning}
{Froning}, C.~S., {Maccarone}, T.~J., {France}, K., {Winter}, L., {Robinson},
  E.~L., {Hynes}, R.~I., \& {Lewis}, F. 2014, ApJ, 780, 48

\bibitem[{{Gallo}(2007)}]{2007Gallob}
{Gallo}, E. 2007, in American Institute of Physics Conference Series, Vol. 924,
  The Multicolored Landscape of Compact Objects and Their Explosive Origins,
  ed. T.~{di Salvo}, G.~L. {Israel}, L.~{Piersant}, L.~{Burderi}, G.~{Matt},
  A.~{Tornambe}, \& M.~T. {Menna}, 715--722

\bibitem[{{Gallo} {et~al.}(2003){Gallo}, {Fender}, \& {Pooley}}]{2003Gallo}
{Gallo}, E., {Fender}, R.~P., \& {Pooley}, G.~G. 2003, MNRAS, 344, 60

\bibitem[{{Gallo} {et~al.}(2014){Gallo}, {Miller-Jones}, {Russell}, {Jonker},
  {Homan}, {Plotkin}, {Markoff}, {Miller}, {Corbel}, \& {Fender}}]{2014Gallo}
{Gallo}, E., {et~al.} 2014, MNRAS, 445, 290

\bibitem[{{Gandhi} {et~al.}(2011){Gandhi}, {Blain}, {Russell}, {Casella},
  {Malzac}, {Corbel}, {D'Avanzo}, {Lewis}, {Markoff}, {Cadolle Bel}, {Goldoni},
  {Wachter}, {Khangulyan}, \& {Mainzer}}]{2011Gandhi}
{Gandhi}, P., {et~al.} 2011, ApJL, 740, L13+

\bibitem[{{Gehrels} {et~al.}(2004){Gehrels}, {Chincarini}, {Giommi}, {Mason},
  {Nousek}, {Wells}, {White}, {Barthelmy}, {Burrows}, {Cominsky}, {Hurley},
  {Marshall}, {M{\'e}sz{\'a}ros}, {Roming}, {Angelini}, {Barbier}, {Belloni},
  {Campana}, {Caraveo}, {Chester}, {Citterio}, {Cline}, {Cropper}, {Cummings},
  {Dean}, {Feigelson}, {Fenimore}, {Frail}, {Fruchter}, {Garmire}, {Gendreau},
  {Ghisellini}, {Greiner}, {Hill}, {Hunsberger}, {Krimm}, {Kulkarni}, {Kumar},
  {Lebrun}, {Lloyd-Ronning}, {Markwardt}, {Mattson}, {Mushotzky}, {Norris},
  {Osborne}, {Paczynski}, {Palmer}, {Park}, {Parsons}, {Paul}, {Rees},
  {Reynolds}, {Rhoads}, {Sasseen}, {Schaefer}, {Short}, {Smale}, {Smith},
  {Stella}, {Tagliaferri}, {Takahashi}, {Tashiro}, {Townsley}, {Tueller},
  {Turner}, {Vietri}, {Voges}, {Ward}, {Willingale}, {Zerbi}, \&
  {Zhang}}]{2004Gehrels}
{Gehrels}, N., {et~al.} 2004, ApJ, 611, 1005

\bibitem[{{Gierli{\'n}ski} {et~al.}(2008){Gierli{\'n}ski}, {Done}, \&
  {Page}}]{2008Gierlinski}
{Gierli{\'n}ski}, M., {Done}, C., \& {Page}, K. 2008, MNRAS, 388, 753

\bibitem[{{Gierli{\'n}ski} {et~al.}(2009){Gierli{\'n}ski}, {Done}, \&
  {Page}}]{2009Gierlinski}
---. 2009, MNRAS, 392, 1106

\bibitem[{{G{\"u}ver} \& {{\"O}zel}(2009)}]{2009Guver}
{G{\"u}ver}, T., \& {{\"O}zel}, F. 2009, MNRAS, 400, 2050

\bibitem[{{Hannikainen} {et~al.}(1998){Hannikainen}, {Hunstead},
  {Campbell-Wilson}, \& {Sood}}]{1998Hannikainen}
{Hannikainen}, D.~C., {Hunstead}, R.~W., {Campbell-Wilson}, D., \& {Sood},
  R.~K. 1998, A\&A, 337, 460

\bibitem[{{Heinz} \& {Sunyaev}(2003)}]{2003Heinz}
{Heinz}, S., \& {Sunyaev}, R.~A. 2003, MNRAS, 343, L59

\bibitem[{{Jenniskens} \& {Desert}(1994)}]{1994Jenniskens}
{Jenniskens}, P., \& {Desert}, F.-X. 1994, A\&AS, 106, 39

\bibitem[{{Jimenez-Garate} {et~al.}(2002){Jimenez-Garate}, {Raymond}, \&
  {Liedahl}}]{2002Jimenez}
{Jimenez-Garate}, M.~A., {Raymond}, J.~C., \& {Liedahl}, D.~A. 2002, ApJ, 581,
  1297

\bibitem[{{La Dous}(1989)}]{1989Ladous}
{La Dous}, C. 1989, A\&A, 211, 131

\bibitem[{{Liu} \& {Meyer-Hofmeister}(2001)}]{2001Liu}
{Liu}, B.~F., \& {Meyer-Hofmeister}, E. 2001, \aap, 372, 386

\bibitem[{{Liu} {et~al.}(2007){Liu}, {Taam}, {Meyer-Hofmeister}, \&
  {Meyer}}]{2007Liu}
{Liu}, B.~F., {Taam}, R.~E., {Meyer-Hofmeister}, E., \& {Meyer}, F. 2007, \apj,
  671, 695

\bibitem[{{Mainzer} {et~al.}(2014){Mainzer}, {Bauer}, {Cutri}, {Grav},
  {Masiero}, {Beck}, {Clarkson}, {Conrow}, {Dailey}, {Eisenhardt}, {Fabinsky},
  {Fajardo-Acosta}, {Fowler}, {Gelino}, {Grillmair}, {Heinrichsen}, {Kendall},
  {Kirkpatrick}, {Liu}, {Masci}, {McCallon}, {Nugent}, {Papin}, {Rice},
  {Royer}, {Ryan}, {Sevilla}, {Sonnett}, {Stevenson}, {Thompson}, {Wheelock},
  {Wiemer}, {Wittman}, {Wright}, \& {Yan}}]{2014Mainzer}
{Mainzer}, A., {et~al.} 2014, ApJ, 792, 30

\bibitem[{{Meyer-Hofmeister} \& {Meyer}(2014)}]{2014Meyerh}
{Meyer-Hofmeister}, E., \& {Meyer}, F. 2014, A\&A, 562, A142

\bibitem[{{Mostafa} {et~al.}(2013){Mostafa}, {Mendez}, {Hiemstra}, {Soleri},
  {Belloni}, {Ibrahim}, \& {Yasein}}]{2013Mostafa}
{Mostafa}, R., {Mendez}, M., {Hiemstra}, B., {Soleri}, P., {Belloni}, T.,
  {Ibrahim}, A.~I., \& {Yasein}, M.~N. 2013, MNRAS, 431, 2341

\bibitem[{{Munari} \& {Zwitter}(1997)}]{1997Munari}
{Munari}, U., \& {Zwitter}, T. 1997, A\&A, 318, 269

\bibitem[{{Narayan} \& {McClintock}(2008)}]{2008Narayan}
{Narayan}, R., \& {McClintock}, J.~E. 2008,  New Astron. Rev., 51, 733

\bibitem[{{Narayan} \& {Yi}(1995)}]{1995Narayan}
{Narayan}, R., \& {Yi}, I. 1995, ApJ, 452, 710

\bibitem[{{Neustroev} {et~al.}(2014){Neustroev}, {Veledina}, {Poutanen},
  {Zharikov}, {Tsygankov}, {Sjoberg}, \& {Kajava}}]{2014Neustroev}
{Neustroev}, V.~V., {Veledina}, A., {Poutanen}, J., {Zharikov}, S.~V.,
  {Tsygankov}, S.~S., {Sjoberg}, G., \& {Kajava}, J.~J.~E. 2014, MNRAS, 445,
  2424

\bibitem[{{Palmer} {et~al.}(2005){Palmer}, {Barthelmey}, {Cummings}, {Gehrels},
  {Krimm}, {Markwardt}, {Sakamoto}, \& {Tueller}}]{2005Palmer}
{Palmer}, D.~M., {Barthelmey}, S.~D., {Cummings}, J.~R., {Gehrels}, N.,
  {Krimm}, H.~A., {Markwardt}, C.~B., {Sakamoto}, T., \& {Tueller}, J. 2005,
  The Astronomer's Telegram, 546, 1

\bibitem[{{Poznanski} {et~al.}(2012){Poznanski}, {Prochaska}, \&
  {Bloom}}]{2012Poznanski}
{Poznanski}, D., {Prochaska}, J.~X., \& {Bloom}, J.~S. 2012, MNRAS, 426, 1465

\bibitem[{{Rahoui} {et~al.}(2010){Rahoui}, {Chaty}, {Rodriguez}, {Fuchs},
  {Mirabel}, \& {Pooley}}]{2010Rahoui}
{Rahoui}, F., {Chaty}, S., {Rodriguez}, J., {Fuchs}, Y., {Mirabel}, I.~F., \&
  {Pooley}, G.~G. 2010, ApJ, 715, 1191

\bibitem[{{Rahoui} {et~al.}(2014){Rahoui}, {Coriat}, \& {Lee}}]{2014Rahouib}
{Rahoui}, F., {Coriat}, M., \& {Lee}, J.~C. 2014, MNRAS, 442, 1610

\bibitem[{{Rahoui} {et~al.}(2011){Rahoui}, {Lee}, {Heinz}, {Hines},
  {Pottschmidt}, {Wilms}, \& {Grinberg}}]{2011Rahoui}
{Rahoui}, F., {Lee}, J.~C., {Heinz}, S., {Hines}, D.~C., {Pottschmidt}, K.,
  {Wilms}, J., \& {Grinberg}, V. 2011, ApJ, 736, 63

\bibitem[{{Rahoui} {et~al.}(2012){Rahoui}, {Coriat}, {Corbel}, {Cadolle Bel},
  {Tomsick}, {Lee}, {Rodriguez}, {Russell}, \& {Migliari}}]{2012Rahoui}
{Rahoui}, F., {et~al.} 2012, MNRAS, 422, 2202

\bibitem[{{Ratti} {et~al.}(2012){Ratti}, {Jonker}, {Miller-Jones}, {Torres},
  {Homan}, {Markoff}, {Tomsick}, {Kaaret}, {Wijnands}, {Gallo}, {{\"O}zel},
  {Steeghs}, \& {Fender}}]{2012Ratti}
{Ratti}, E.~M., {et~al.} 2012, \mnras, 423, 2656

\bibitem[{{Reis} {et~al.}(2009){Reis}, {Fabian}, {Ross}, \&
  {Miller}}]{2009Reis}
{Reis}, R.~C., {Fabian}, A.~C., {Ross}, R.~R., \& {Miller}, J.~M. 2009, MNRAS,
  395, 1257

\bibitem[{{Reynolds} {et~al.}(2010){Reynolds}, {Miller}, {Homan}, \&
  {Miniutti}}]{2010Reynolds}
{Reynolds}, M.~T., {Miller}, J.~M., {Homan}, J., \& {Miniutti}, G. 2010, ApJ,
  709, 358

\bibitem[{{Roming} {et~al.}(2005){Roming}, {Kennedy}, {Mason}, {Nousek}, {Ahr},
  {Bingham}, {Broos}, {Carter}, {Hancock}, {Huckle}, {Hunsberger}, {Kawakami},
  {Killough}, {Koch}, {McLelland}, {Smith}, {Smith}, {Soto}, {Boyd},
  {Breeveld}, {Holland}, {Ivanushkina}, {Pryzby}, {Still}, \&
  {Stock}}]{2005Roming}
{Roming}, P.~W.~A., {et~al.} 2005, \ssr, 120, 95

\bibitem[{{Russell} {et~al.}(2006){Russell}, {Fender}, {Hynes}, {Brocksopp},
  {Homan}, {Jonker}, \& {Buxton}}]{2006Russell}
{Russell}, D.~M., {Fender}, R.~P., {Hynes}, R.~I., {Brocksopp}, C., {Homan},
  J., {Jonker}, P.~G., \& {Buxton}, M.~M. 2006, MNRAS, 371, 1334

\bibitem[{{Russell} {et~al.}(2010){Russell}, {Maitra}, {Dunn}, \&
  {Markoff}}]{2010Russell}
{Russell}, D.~M., {Maitra}, D., {Dunn}, R.~J.~H., \& {Markoff}, S. 2010, MNRAS,
  405, 1759

\bibitem[{{Russell} {et~al.}(2013){Russell}, {Markoff}, {Casella}, {Cantrell},
  {Chatterjee}, {Fender}, {Gallo}, {Gandhi}, {Homan}, {Maitra}, {Miller-Jones},
  {O'Brien}, \& {Shahbaz}}]{2013Russell}
{Russell}, D.~M., {et~al.} 2013, MNRAS, 429, 815

\bibitem[{{Sault} {et~al.}(1995){Sault}, {Teuben}, \& {Wright}}]{1995Sault}
{Sault}, R.~J., {Teuben}, P.~J., \& {Wright}, M.~C.~H. 1995, in Astronomical
  Society of the Pacific Conference Series, Vol.~77, Astronomical Data Analysis
  Software and Systems IV, ed. R.~A. {Shaw}, H.~E. {Payne}, \& J.~J.~E.
  {Hayes}, 433

\bibitem[{{Shakura} \& {Syunyaev}(1973)}]{1973Shakura}
{Shakura}, N.~I., \& {Syunyaev}, R.~A. 1973, A\&A, 24, 337

\bibitem[{{Shaw} {et~al.}(2013){Shaw}, {Charles}, {Bird}, {Cornelisse},
  {Casares}, {Lewis}, {Mu{\~n}oz-Darias}, {Russell}, \& {Zurita}}]{2013Shaw}
{Shaw}, A.~W., {et~al.} 2013, MNRAS, 433, 740

\bibitem[{{Soleri} \& {Fender}(2011)}]{2011Soleri}
{Soleri}, P., \& {Fender}, R. 2011, MNRAS, 413, 2269

\bibitem[{{Soria} {et~al.}(2000){Soria}, {Wu}, \& {Hunstead}}]{2000Soria}
{Soria}, R., {Wu}, K., \& {Hunstead}, R.~W. 2000, ApJ, 539, 445

\bibitem[{{Titarchuk}(1994)}]{1994Titarchuk}
{Titarchuk}, L. 1994, ApJ, 434, 570

\bibitem[{{Tomsick} {et~al.}(2015){Tomsick}, {Rahoui}, {Kolehmainen},
  {Miller-Jones}, {Fuerst}, {Yamaoka}, {Akitaya}, {Corbel}, {Coriat}, {Done},
  {Gandhi}, {Harrison}, {Huang}, {Kaaret}, {Kalemci}, {Kanda}, {Migliari},
  {Miller}, {Moritani}, {Stern}, {Uemura}, \& {Urata}}]{2015Tomsick}
{Tomsick}, J.~A., {et~al.} 2015, ApJ, in press, astro-ph 1506.06780

\bibitem[{{Torres} {et~al.}(2005){Torres}, {Steeghs}, {Blake}, {Jonker},
  {Garcia}, {McClintock}, {Miller}, {Zhao}, {Calkins}, {Berlind}, {Falco},
  {Bloom}, {Callanan}, \& {Rodriguez-Gil}}]{2005Torres}
{Torres}, M.~A.~P., {et~al.} 2005, The Astronomer's Telegram, 566, 1

\bibitem[{{Vernet} {et~al.}(2011){Vernet}, {Dekker}, {D'Odorico}, {Kaper},
  {Kjaergaard}, {Hammer}, {Randich}, {Zerbi}, {Groot}, {Hjorth}, {Guinouard},
  {Navarro}, {Adolfse}, {Albers}, {Amans}, {Andersen}, {Andersen}, {Binetruy},
  {Bristow}, {Castillo}, {Chemla}, {Christensen}, {Conconi}, {Conzelmann},
  {Dam}, {de Caprio}, {de Ugarte Postigo}, {Delabre}, {di Marcantonio},
  {Downing}, {Elswijk}, {Finger}, {Fischer}, {Flores}, {Fran{\c c}ois},
  {Goldoni}, {Guglielmi}, {Haigron}, {Hanenburg}, {Hendriks}, {Horrobin},
  {Horville}, {Jessen}, {Kerber}, {Kern}, {Kiekebusch}, {Kleszcz}, {Klougart},
  {Kragt}, {Larsen}, {Lizon}, {Lucuix}, {Mainieri}, {Manuputy}, {Martayan},
  {Mason}, {Mazzoleni}, {Michaelsen}, {Modigliani}, {Moehler}, {M{\o}ller},
  {Norup S{\o}rensen}, {N{\o}rregaard}, {P{\'e}roux}, {Patat}, {Pena}, {Pragt},
  {Reinero}, {Rigal}, {Riva}, {Roelfsema}, {Royer}, {Sacco}, {Santin},
  {Schoenmaker}, {Spano}, {Sweers}, {Ter Horst}, {Tintori}, {Tromp}, {van
  Dael}, {van der Vliet}, {Venema}, {Vidali}, {Vinther}, {Vola}, {Winters},
  {Wistisen}, {Wulterkens}, \& {Zacchei}}]{2011Vernet}
{Vernet}, J., {et~al.} 2011, A\&A, 536, A105

\bibitem[{{Wilms} {et~al.}(2000){Wilms}, {Allen}, \& {McCray}}]{2000Wilms}
{Wilms}, J., {Allen}, A., \& {McCray}, R. 2000, ApJ, 542, 914

\bibitem[{{Wilson} {et~al.}(2011){Wilson}, {Ferris}, {Axtens}, {Brown},
  {Davis}, {Hampson}, {Leach}, {Roberts}, {Saunders}, {Koribalski}, {Caswell},
  {Lenc}, {Stevens}, {Voronkov}, {Wieringa}, {Brooks}, {Edwards}, {Ekers},
  {Emonts}, {Hindson}, {Johnston}, {Maddison}, {Mahony}, {Malu}, {Massardi},
  {Mao}, {McConnell}, {Norris}, {Schnitzeler}, {Subrahmanyan}, {Urquhart},
  {Thompson}, \& {Wark}}]{2011Wilson}
{Wilson}, W.~E., {et~al.} 2011, MNRAS, 416, 832

\bibitem[{{Wright} {et~al.}(2010){Wright}, {Eisenhardt}, {Mainzer}, {Ressler},
  {Cutri}, {Jarrett}, {Kirkpatrick}, {Padgett}, {McMillan}, {Skrutskie},
  {Stanford}, {Cohen}, {Walker}, {Mather}, {Leisawitz}, {Gautier}, {McLean},
  {Benford}, {Lonsdale}, {Blain}, {Mendez}, {Irace}, {Duval}, {Liu}, {Royer},
  {Heinrichsen}, {Howard}, {Shannon}, {Kendall}, {Walsh}, {Larsen}, {Cardon},
  {Schick}, {Schwalm}, {Abid}, {Fabinsky}, {Naes}, \& {Tsai}}]{2010Wright}
{Wright}, E.~L., {et~al.} 2010, AJ, 140, 1868

\bibitem[{{Wu} {et~al.}(2001){Wu}, {Soria}, {Hunstead}, \& {Johnston}}]{2001Wu}
{Wu}, K., {Soria}, R., {Hunstead}, R.~W., \& {Johnston}, H.~M. 2001, MNRAS,
  320, 177

\bibitem[{{Zhang} {et~al.}(2010){Zhang}, {Yuan}, \& {Chaty}}]{2010Zhang}
{Zhang}, H., {Yuan}, F., \& {Chaty}, S. 2010, ApJ, 717, 929

\bibitem[{{Zurita} {et~al.}(2008){Zurita}, {Durant}, {Torres}, {Shahbaz},
  {Casares}, \& {Steeghs}}]{2008Zuritab}
{Zurita}, C., {Durant}, M., {Torres}, M.~A.~P., {Shahbaz}, T., {Casares}, J.,
  \& {Steeghs}, D. 2008, ApJ, 681, 1458

\end{thebibliography}

\end{document}